\documentclass[conference,compsoc]{IEEEtran}

\ifCLASSOPTIONcompsoc
  \usepackage[nocompress]{cite}
\else
  \usepackage{cite}
\fi

\usepackage[T1]{fontenc}
\usepackage[utf8]{inputenc}
\usepackage{url}
\usepackage{xspace}

\usepackage{adjustbox}
\usepackage{amsmath}

\usepackage{amssymb}

\usepackage{graphicx}
\usepackage{booktabs}
\usepackage{array}
\usepackage{multirow}
\usepackage{multicol}
\usepackage{makecell}
\usepackage{diagbox}
\usepackage{colortbl}
\usepackage{threeparttable}
\usepackage{longtable}
\usepackage{rotating}
\usepackage{lscape}
\usepackage{wrapfig}

\ifCLASSOPTIONcompsoc
  \usepackage[caption=false,font=footnotesize,labelfont=sf,textfont=sf]{subfig}
\else
  \usepackage[caption=false,font=footnotesize]{subfig}
\fi

\usepackage{algorithmic}
\usepackage[ruled,vlined,lined,commentsnumbered]{algorithm2e}

\usepackage{listings}
\usepackage{moreverb}
\usepackage{soul}
\usepackage{color}
\usepackage{tikz}
\usepackage[most]{tcolorbox}
\usepackage{pgfplots}
\pgfplotsset{compat=1.18}
\usepackage{nicematrix}
\usepackage[hidelinks]{hyperref}

\usepackage{enumitem}
\usepackage{csvsimple}
\usepackage{fontawesome}
\usepackage{dialogue}
\usepackage{fancybox}
\usepackage{marvosym}
\usepackage{pifont}
\usepackage{balance}

\newcommand{\find}[1]{
\begin{tcolorbox}[
  tile,
  fontupper=\small,
  size=fbox,
  boxsep=2.2mm,
  boxrule=0pt,
  top=0pt,
  bottom=0pt,
  borderline west={0.5mm}{0pt}{black!50!white},
  colback=black!5!white
]
#1
\end{tcolorbox}
}

\newcommand{\toolname}{{\sc BinMirror}\xspace}
\newcommand{\bftoolname}{\textbf{\textsc{BinMirror}}\xspace}
\newcommand{\bfittoolname}{\textbf{\textsc{\textit{BinMirror}}}\xspace}

\begin{document}

\title{Behavior Specification-Guided Program Synthesis for Binary Deobfuscation}

\author{
\IEEEauthorblockN{
\mbox{Kangchen Zhu\IEEEauthorrefmark{1}\IEEEauthorrefmark{2}},
\quad
\mbox{Shangwen Wang\IEEEauthorrefmark{1}\IEEEauthorrefmark{3}},
\quad
\mbox{Zhiliang Tian\IEEEauthorrefmark{1}},
\quad
\mbox{Zhouyang Jia\IEEEauthorrefmark{1}}
\\[0.6ex]
\mbox{Xiaoling Li\IEEEauthorrefmark{1}},
\quad
\mbox{Jun Ma\IEEEauthorrefmark{1}},
\quad
\mbox{Jie Yu\IEEEauthorrefmark{1}},
\quad
\mbox{Xiaoguang Mao\IEEEauthorrefmark{1}\IEEEauthorrefmark{2}}
}

\IEEEauthorblockA{
\IEEEauthorrefmark{1}
College of Computer Science and Technology,
National University of Defense Technology,
Changsha, China
}

\IEEEauthorblockA{
\IEEEauthorrefmark{2}
Also with the State Key Laboratory of Complex and Critical Software Environment
}

\IEEEauthorblockA{
\IEEEauthorrefmark{3}
Corresponding authors
}
}

\maketitle

\begin{abstract}
Deobfuscation is critical for reverse engineering and security analysis, as it restores readability and analyzability to obfuscated code. 
However, existing research primarily targets source-code deobfuscation, leaving binary-level deobfuscation, a more critical task given the unavailability of source code in real-world scenarios, largely underexplored. 
Current binary deobfuscation methods typically decompile binaries into pseudocode before applying structural transformations. 
However, this decompilation-based paradigm suffers from the loss of high-level semantics during compilation, such as precise types and source-level structures, often producing low-quality code and making it difficult to validate whether the recovered code preserves the intended runtime behavior of the original program. 
To address these limitations, we propose a paradigm shift from structural transformation to behavior-driven synthesis. 
Our core insight is that although obfuscation distorts a program's internal structure, semantics-preserving transformations must retain its observable execution behavior. 
Building on this insight, we introduce \bftoolname, an approach that reframes binary deobfuscation as a behavior-specification-guided program synthesis task. 
By treating dynamic execution traces and interaction snapshots as behavioral specifications, \bftoolname synthesizes high-quality source code that is validated against observed runtime behaviors extracted from heavily obfuscated binaries. 
Extensive evaluations on 1.5 million synthetic obfuscated binaries show that \bftoolname significantly outperforms state-of-the-art baselines, achieving a 74.5\% unit-test Pass@1 under extreme obfuscation. Furthermore, integrating \bftoolname into a downstream malware detection pipeline improves the detection of obfuscated in-the-wild malware, increasing accuracy by 33.3\% and F1-score by 37.1\% compared with direct binary analysis. 
These results demonstrate the practical utility of \bftoolname for restoring semantic clarity in real-world security analysis.
\end{abstract}


\IEEEpeerreviewmaketitle

\section{Introduction}
Code obfuscation is a widely adopted software protection mechanism that applies semantics-preserving transformations to render programs structurally opaque while preserving their intended functionality~\cite{mariano2024control}. Common transformations include dead code insertion~\cite{barria2016obfuscation}, identifier renaming~\cite{chan2004advanced}, instruction substitution~\cite{zhou2007information}, and control-flow flattening~\cite{johansson2017lightweight}. While these techniques help protect intellectual property~\cite{schrittwieser2016protecting, OLLVM_Theory}, they are also frequently abused by attackers to hide malicious behaviors and evade detection~\cite{you2010malware}. Therefore, deobfuscation is essential for reverse engineering and downstream security tasks such as malware analysis and software plagiarism detection.

Existing deobfuscation techniques are largely designed for source-code settings~\cite{Bardin2017ValueAnalysis, dong2022cadecff, li2022generic, choi2024chatdeob, mariano2024control, chen2025jsdeobsbench}. These methods benefit from source-level information such as precise types, meaningful identifiers, and structured syntax, which guide pattern matching, program transformation, or learning-based rewriting. However, such assumptions rarely hold in practical security scenarios, where analysts often face stripped binary executables rather than source code~\cite{tkachenko2025deconstructing, liu2025llm2, david2020qsynth}. A common workaround is to first decompile the obfuscated binary into C-like pseudocode and then apply source-level deobfuscation tools~\cite{mariano2024control, choi2024chatdeob}. Unfortunately, compilation and decompilation can remove or mis-recover high-level semantics, including variable types, source-level control structures, and contextual metadata~\cite{ye2023cp}. As a result, decompilation-based pipelines may produce misleading pseudocode and make it difficult to check whether a reconstructed representation preserves observed runtime behavior.

To overcome these limitations, we propose a paradigm shift: bypassing the static pattern matching on flawed decompiled pseudocode in favor of analyzing dynamic execution behavior.
Our core insight is that \textit{while semantics-preserving obfuscation distorts a program's internal structure and syntax, it must preserve the program's observable behavior on executed paths}~\cite{jain2022indistinguishability, tofighi2018dose}.
Specifically, for a given execution, the program's externally observable interactions and side-effects should remain consistent~\cite{schrittwieser2013covert, banescu2016code}.
Here, we define execution behavior as the set of characteristics observed during runtime~\cite{baralis2002compile}, and side-effects as state changes beyond simple value computation, such as modifying non-local variables, updating memory through references, or performing I/O operations~\cite{klein2009sel4}.
Therefore, rather than treating decompiled pseudo as a reliable semantic substrate, we use observed runtime effects as an empirical behavioral anchor for guiding deobfuscation.

Building upon the above observation, we propose \toolname, a novel end-to-end approach that reframes binary deobfuscation as a behavior-specification-guided program synthesis task. This formulation does not claim that program synthesis is new in general; instead, its novelty lies in applying behavior-constrained synthesis to security-oriented binary deobfuscation, where static structure and decompiled pseudocode are often unreliable. Recognizing that Large Language Models (LLMs) have set new benchmarks in semantic inference and code synthesis~\cite{choi2024chatdeob, joel2024survey, wang2023review}, \toolname integrates an LLM as its core synthesis engine to translate low-level behavioral specifications, extracted from dynamic execution traces and interaction snapshots, into readable source-level reconstructions that are consistent with observed behavior.
Specifically, \toolname operates in four sequential stages:
(1) The \textit{Behavior Specification Capture } stage observes the program's dynamic runtime interactions and captures critical external events to construct an observed behavioral specification.
By combining syscall-guided greybox fuzzing with dynamic binary instrumentation, it records execution context snapshots at key interaction points.
(2) The \textit{Specification-Driven Noise Filtering} stage extracts the instruction sequence responsible for the observed side-effects.
It uses trace-enhanced backward data-flow slicing to filter out executed but behavior-irrelevant trace instructions, such as dispatcher updates, opaque-predicate computations, and MBA junk computations that do not contribute to the monitored behavior.
(3) The \textit{Specification-Guided Code Synthesis} stage reconstructs readable source-level code under the constraints imposed by the captured behavioral specifications.
Leveraging a large language model as a semantic lifter, this stage constrains the code generation using captured input/output states and purified instruction slices, reducing unconstrained hallucination during binary-to-source lifting.
Finally, (4) the \textit{Test-Based Behavioral Validation} stage executes the candidate code within a context-aware test harness to perform differential testing against the original binary on observed executions.
Any behavioral discrepancies are fed back to the LLM through a closed-loop iterative refinement mechanism, enabling the system to iteratively improve the candidate code until it matches the captured behavioral specifications or reaches a predefined iteration bound.

To comprehensively evaluate the effectiveness and practical utility of \toolname, we conducted extensive experiments.
First, we constructed a massive synthetic benchmark, which consists of 1.5 million unique obfuscated binaries.
We generate this dataset by applying four industry-standard obfuscators to obfuscate the collected source programs across four instruction architectures (x86, x64, ARM, MIPS) and four compiler optimization levels (O0-O3).
We then evaluate the reconstructed code using Unit Test Pass Rate (Pass@1), which measures whether the compiled reconstruction passes the available source-level tests.
Our results demonstrate that \toolname significantly outperforms state-of-the-art (SOTA) baselines, achieving a 74.5\% unit-test Pass@1 under extreme obfuscation scenarios.
To further understand when \toolname and existing tools succeed or fail, we also conduct a sample-level overlap and failure analysis, showing that \toolname is complementary to static, dynamic/symbolic, and learning-based baselines rather than uniformly dominating them.
Furthermore, to assess the quality and readability of the source-level reconstructions produced from multiple obfuscation levels, we measured the CodeBLEU~\cite{choi2024chatdeob, patsakis2024assessing} and cyclomatic complexity~\cite{liu2025llm2}.
The results show that \toolname yields clean, readable code, achieving an average of 68.4\% CodeBLEU score and 76.2\% reduction in cyclomatic complexity under extreme obfuscation.
Finally, we demonstrated the downstream utility of our approach by integrating \toolname into a CFG-based Linux malware detector.
The integration empowered existing detection engines to detect obfuscated malware with higher accuracy (+33.3\%) and F1-scores (+37.1\%), confirming that \toolname is effective in facilitating security analysis tasks.
Notably, these validation results are bounded by the available tests and observed executions; \toolname does not prove semantic equivalence over all possible inputs.

In summary, our contributions are as follows:
\begin{itemize}[left=0pt, nosep]

\item \textbf{Paradigm Shift}: We introduce a binary deobfuscation paradigm that shifts from traditional syntax-driven structural transformations to behavior-specification-guided program synthesis.

\item \textbf{Approach}: We propose \toolname, an end-to-end framework that combines syscall-guided behavior capture, trace-enhanced slicing, LLM-based synthesis, and differential testing to generate source-level reconstructions validated against observed runtime behaviors.

\item \textbf{Extensive Evaluation}: We evaluate \toolname on 1.5 million synthetic obfuscated binaries and real-world obfuscated malware, demonstrating strong test-based recovery effectiveness, readability improvement, and downstream malware-detection utility\footnote{The code and data used in our evaluation are available at \url{https://zenodo.org/records/19067406}.}.

\end{itemize}

\section{Background and Related Work}
\label{sec:background_related_work}

Binary deobfuscation aims to recover readable and analyzable code from obfuscated binaries. 
In this paper, we focus on stripped binaries for which source-level information, such as variable names, precise types, and structured control flow, is unavailable or unreliable after compilation and decompilation. 
Our goal is to synthesize readable source code that is consistent with observed runtime behaviors and available tests. 
This validation is test-based and observation-bounded, rather than a proof of semantic equivalence over all possible inputs.

\begin{figure*}[!t]
  \centering
  \includegraphics[width=0.98\linewidth]{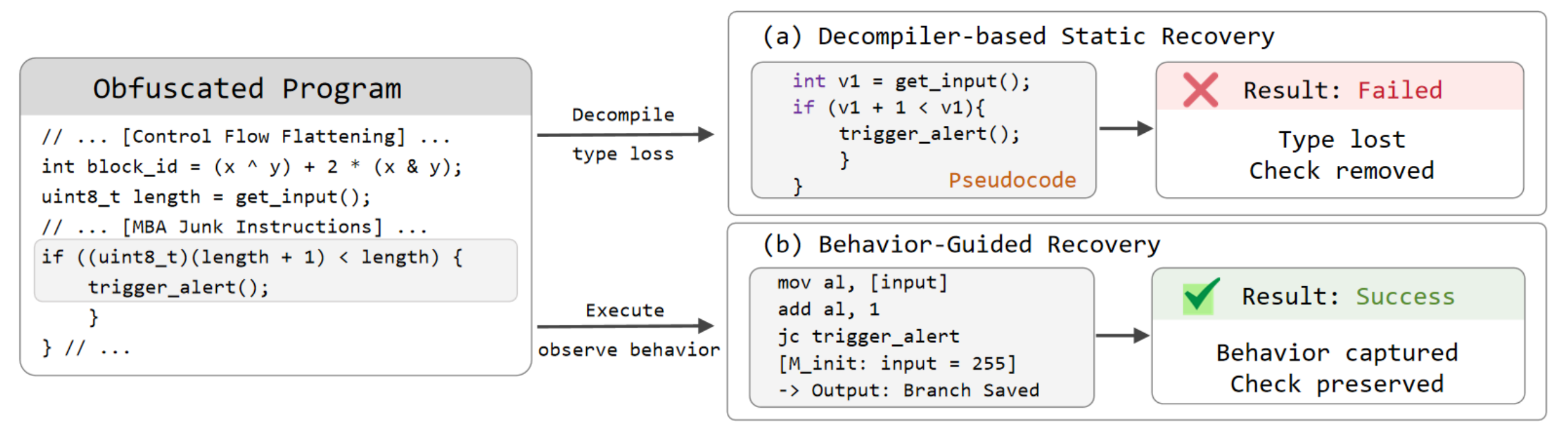} 
  \caption{A motivation example. 
  (a) Type information such as \texttt{uint8\_t} may be lost or mis-recovered during decompilation, causing a static deobfuscator or simplifier to incorrectly remove an overflow check from the recovered pseudocode. 
  (b) By capturing execution-derived behavioral specifications, our approach constrains synthesis using observed runtime behavior and validates the recovered code against the captured behavior.}
  \label{fig:motivation_example}
\end{figure*}

\noindent\textbf{\textit{Static, symbolic, and dynamic deobfuscation.}}
Existing binary deobfuscation techniques commonly rely on static analysis, solver-aided reasoning, or dynamic trace analysis~\cite{shirazi2019analysis}. 
Static tools inspect code structure without execution and apply pattern matching, algebraic simplification, data-flow analysis, or SMT-based reasoning to simplify obfuscation primitives~\cite{idapro,dong2022cadecff,d810github,li2024x}. 
For instance, D810~\cite{d810github} integrates into the Hex-Rays~\cite{idapro} microcode pipeline to perform backward variable tracking and pattern-based simplification. 
Dynamic and trace-based methods execute programs to observe concrete values, executed paths, and runtime side effects~\cite{nethercote2007valgrind,david2020qsynth,mariano2024control,blazytko2017syntia,menguy2021search}. 
Representative systems such as QSynth~\cite{david2020qsynth} and Syntia~\cite{blazytko2017syntia} combine dynamic sampling, symbolic reasoning, or synthesis to simplify obfuscated expressions. 
However, these methods are often optimized for localized simplification rather than end-to-end source recovery. 
They can be limited by predefined rules, state explosion, path coverage, and the cost of symbolic or enumerative search, especially when heavy control-flow flattening, bogus branches, MBA transformations, and decompiler-induced type loss make the static representation fragmented or misleading.

\noindent\textbf{\textit{Learning-based and LLM-based deobfuscation.}}
Recent work increasingly explores learning-based deobfuscation, with LLMs showing strong potential for code understanding and rewriting~\cite{choi2024chatdeob,chen2025jsdeobsbench,patsakis2024assessing,liu2025llm2}. 
Prior studies have evaluated fine-tuned or prompt-based models for source-code deobfuscation, JavaScript deobfuscation, and malicious-code simplification~\cite{beste2025exploring,chen2025jsdeobsbench,patsakis2024assessing}. 
For binary-level scenarios, a prevalent pipeline first lifts an obfuscated binary into C-like pseudocode or assembly and then applies source-level optimization or LLM-based rewriting~\cite{mohseni2025can,choi2024chatdeob,mariano2024control,liu2025llm2}. 
For example, CHISEL~\cite{mohseni2025can} performs trace-informed compositional synthesis over decompiled pseudocode, while ChatDEOB~\cite{choi2024chatdeob} uses fine-tuned models after preliminary decompilation. 
Although effective in many cases, these pipelines remain sensitive to the semantic gap introduced by compilation and decompilation: precise types, identifiers, and source-level structure may be lost or mis-recovered, causing source-oriented models to infer incorrect logic or remove security-relevant behavior. 
\toolname addresses this limitation by grounding synthesis in dynamic execution traces and interaction snapshots rather than relying solely on decompiled pseudocode.

\noindent\textbf{\textit{Execution-derived specifications and program synthesis.}}
Our work is also related to fuzzing, execution-based specification mining, and program synthesis. 
Greybox fuzzing generates inputs using lightweight feedback such as coverage~\cite{AFL,AFL++,bohme2016coverage,rawat2017vuzzer}, while specification mining extracts behavioral facts or likely invariants from program executions~\cite{ernst2007daikon}. 
These techniques provide empirical evidence about program behavior, but their results are inherently bounded by observed executions: if an input does not trigger a path, the corresponding behavior cannot be captured. 
Program synthesis has also been explored for decompilation and structure recovery, such as recovering loops from hardware netlists or synthesizing structured programs from domain-specific intermediate representations~\cite{sisco2023loop,nandi2020synthesizing}. 
\toolname is inspired by these directions but targets a different setting: security-oriented deobfuscation of heavily obfuscated stripped binaries, where source-level types, identifiers, and reliable structured pseudocode are often unavailable. 
Instead of recovering structure from explicit domain-specific semantics, \toolname extracts behavioral specifications from dynamic execution, filters trace-level noise, guides LLM-based source-code synthesis, and validates candidate code through differential testing on observed executions. 
Thus, \toolname complements prior work by treating observed runtime behavior as a specification for synthesizing and test-validating readable source code from obfuscated binaries.

\toolname does not claim novelty in dynamic tracing, slicing, synthesis, or LLM inference alone.
Its contribution is a behavior-specification-guided reconstruction pipeline that turns runtime behavior into state-transfer constraints, extracts behavior-relevant binary slices, and synthesizes compilable C code validated on held-out executions, distinguishing it from expression-level synthesis and decompiler+LLM pipelines that mainly rely on recovered pseudocode.

\section{Motivation Example}

\begin{figure*}[!t]
  \centering
  \includegraphics[width=0.85\linewidth]{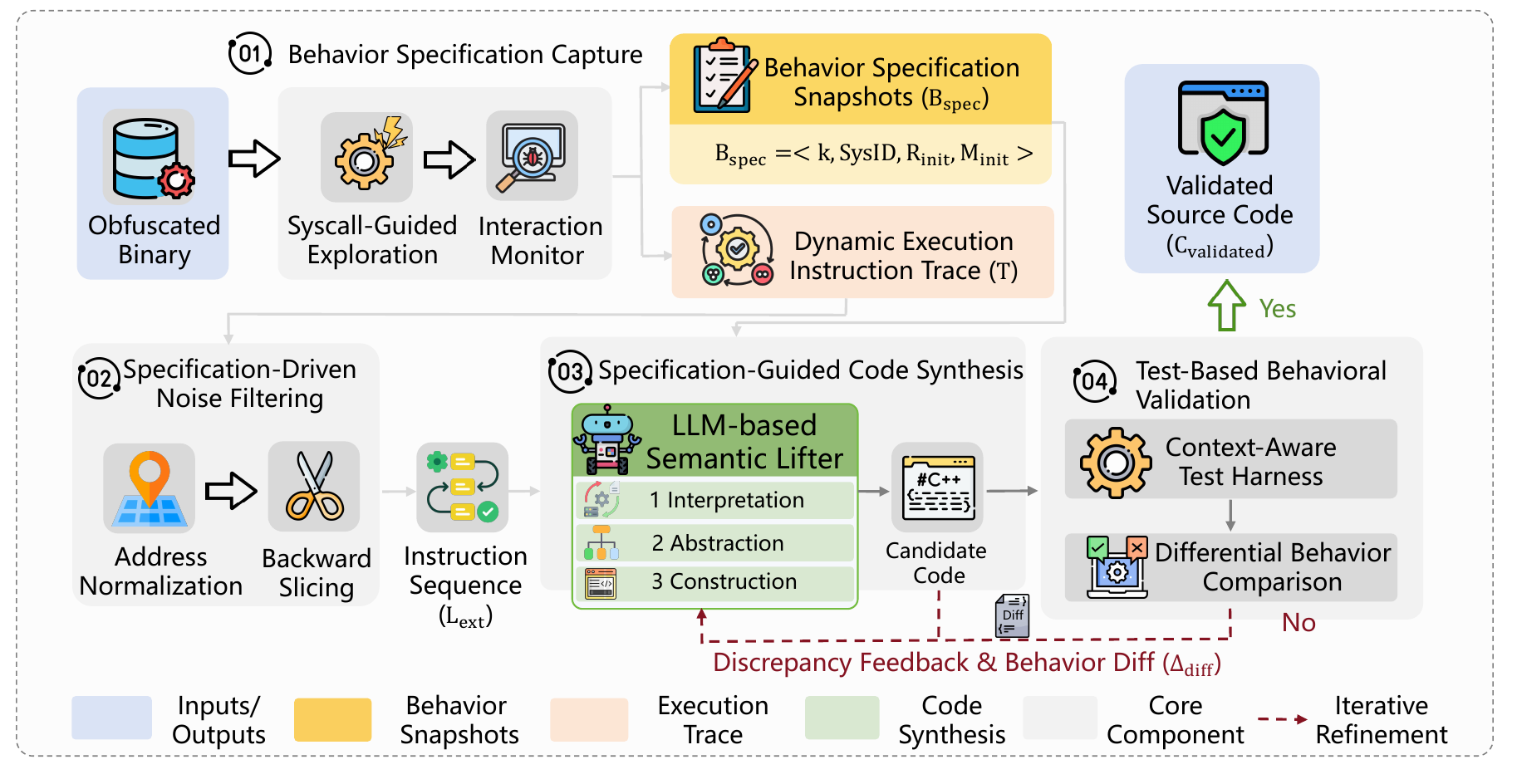} 
  \caption{Overview of \toolname, a four-stage behavior-specification-guided pipeline that captures observed runtime behavior, filters trace-level noise, synthesizes readable source code, and validates candidate programs through differential testing on observed executions.}
  \label{fig:overview}
\end{figure*}

Many current deobfuscation pipelines lift obfuscated binaries into C-like pseudocode and subsequently apply source-level transformations~\cite{mohseni2025can, choi2024chatdeob, mariano2024control, liu2025llm2}. 
However, this approach faces a critical limitation: compilation can remove or obscure high-level semantics, such as function names, source-level control structures, and precise variable types. 
Because the resulting pseudocode may lack this essential context, source-level tools can struggle to accurately interpret the program's logic. 
Hence, this decompilation-based strategy may yield low-quality code or even remove security-relevant behavior when the decompiled representation is semantically misleading.

To illustrate this limitation, we use a standard integer-overflow check as a motivating example.
As shown on the left side of Figure~\ref{fig:motivation_example}, the source code is designed to detect an overflow when incrementing an 8-bit unsigned integer (\texttt{uint8\_t}).
Because a \texttt{uint8\_t} occupies one byte, its valid range is 0 to 255.
When the variable reaches its maximum value of 255, adding 1 mathematically yields 256 (\texttt{100000000} in binary).
However, under the 8-bit type constraint, the value wraps around to 0 (\texttt{00000000} in binary).
Consequently, the condition \texttt{0 < 255} evaluates to true, correctly invoking \texttt{trigger\_alert()}.

During compilation and decompilation, however, the precise \texttt{uint8\_t} type may be lost or mis-recovered, depending on the compiler, optimization level, and decompiler heuristics.
As shown in the upper right part of Figure~\ref{fig:motivation_example}, the decompiler may reconstruct the variable from a generic CPU register and promote it to a wider signed integer type, such as \texttt{int v1}.
Under this recovered type, the expression \texttt{v1 + 1 < v1} no longer captures the original 8-bit wraparound behavior and appears mathematically impossible in ordinary signed-integer arithmetic.
When a static deobfuscator or simplifier processes this misleading pseudocode, it may conclude that the condition is always false and remove the branch as unreachable.
As a result, the decompilation-based pipeline can eliminate a critical security check, even though the original binary still contains runtime behavior that triggers the alert.

To overcome this limitation, our intuition is to shift the focus from transforming fragmented static syntax to analyzing observed dynamic execution behavior. 
For semantics-preserving obfuscation, although internal structure and syntax may be distorted, the externally observable behavior on executed paths should remain consistent.
As shown in the lower right part of Figure~\ref{fig:motivation_example}, our approach leverages this behavioral consistency by capturing concrete runtime states. 
By executing the binary with an overflow-triggering input, e.g., \texttt{255}, we record the initial memory state ($M_{init}$: \texttt{length = 255}) and the observed branch outcome to form an execution-derived behavioral specification.
These constraints provide factual runtime evidence for program synthesis and prevent the synthesizer from relying solely on flawed algebraic assumptions derived from decompiled pseudocode.
Consequently, the synthesized code can preserve the observed wraparound behavior, for example, by generating a type-aware condition such as \texttt{if ((uint8\_t)(length + 1) < (uint8\_t)length)}.

This example highlights a real failure mode of existing decompilation-based deobfuscation pipelines and motivates our proposed solution: prioritizing observed runtime behavior over potentially misleading static syntax.
Our idea shifts the paradigm from syntax-driven structural transformation to behavior-specification-guided program synthesis.
By capturing dynamic execution states and validating synthesized code against the observed behavior, we reframe binary deobfuscation as a behavior-constrained synthesis task.

\section{Approach}
\label{sec:method}

\subsection{Overview}
\label{sec:overview}

Figure~\ref{fig:overview} illustrates the overview of \toolname. 
\toolname follows a four-stage pipeline that transforms observed runtime behavior into behavior-constrained source-code synthesis and test-based validation.
\begin{itemize}[leftmargin=*, nosep]
    \item \textbf{\textit{Phase I: Behavior Specification Capture.}} 
    This stage observes dynamic runtime interactions and captures critical external events as execution-derived behavioral specifications. 
    It combines syscall-guided greybox exploration with dynamic binary instrumentation (DBI) to record execution contexts at key interaction points.
    
    \item \textbf{\textit{Phase II: Specification-Driven Noise Filtering.}} 
    This stage extracts the behavior-relevant instruction sequence from the dynamic trace. 
    Starting from monitored interaction points, it applies trace-enhanced backward slicing to prune executed but behavior-irrelevant instructions, such as dispatcher updates, opaque-predicate computations, and MBA junk instructions.
    
    \item \textbf{\textit{Phase III: Specification-Guided Program Synthesis.}} 
    This stage synthesizes readable source code under the constraints of the captured behavioral specifications. 
    An LLM-based semantic lifter translates the purified instruction slice and its behavior snapshot into candidate code.
    
    \item \textbf{\textit{Phase IV: Test-Based Behavioral Validation.}} 
    This stage executes the candidate code in a context-aware test harness and compares it with the original binary on observed executions. 
    Behavioral discrepancies are fed back to the LLM through closed-loop refinement until the candidate matches the captured specifications or reaches a predefined iteration bound.
\end{itemize}

\subsection{Behavior Specification Capture}
\label{sec:phase1}

This phase captures execution-derived behavioral specifications from the obfuscated binary $\mathcal{B}_{obf}$. 
We focus on externally observable interaction behaviors, such as file I/O and network communication, because semantics-preserving obfuscation may distort internal code structures but should preserve such behaviors on executed paths.
These interactions provide empirical runtime anchors for deobfuscation.

The input is $\mathcal{B}_{obf}$ and an initial seed corpus $\mathcal{I}$.
The output is a set of dynamic instruction traces $\mathcal{T}={T^{(1)},T^{(2)},\dots,T^{(m)}}$ paired with behavior snapshots $\mathcal{B}_{SPEC}={B_{spec}^{(1)},B_{spec}^{(2)},\dots,B_{spec}^{(m)}}$.
Each snapshot is:
\begin{equation}
B_{spec}=\langle k,\text{SysID},\mathcal{R}_{init},\mathcal{M}_{init}\rangle,
\end{equation}
where $k$ is the trace index of the monitored interaction, $\text{SysID}$ is the system call identifier, $\mathcal{R}_{init}$ is the CPU register state, and $\mathcal{M}_{init}$ is the memory regions of the interaction.

\noindent\textbf{\textit{Syscall-Guided Exploration.}}
Traditional fuzzing typically optimizes basic-block coverage ($\text{Cov}_{bb}$)~\cite{AFL, AFL++}, but this signal can be distorted by bogus control flows, flattened dispatchers, and opaque predicates in heavily obfuscated binaries.
To reduce this distortion, \toolname adopts syscall-guided greybox exploration~\cite{xiao2025robust}, prioritizing inputs that trigger new externally observable interaction behaviors rather than merely increasing coverage over artificial control-flow structures. \toolname does not require source-level type information to bootstrap exploration.
If sample inputs are available, they are used as seeds; otherwise, \toolname starts from empty inputs and random byte streams.
During execution, the monitor observes input-related syscalls, such as \texttt{sys\_read}, \texttt{sys\_recv}, \texttt{sys\_recvfrom}, and file-backed input operations, to infer input channels and mutate the corresponding byte streams.
The feedback signal is the novelty of interaction behavior instead of source-level coverage or type information.

We model an execution's observable interaction as a syscall sequence $S_{sys}=\langle s_1,s_2,\dots,s_n\rangle$, where $s\in\mathcal{S}_{syscall}$.
The fuzzer maintains a global $\text{History}$ of observed syscall sequences and preserves an input $i\in\mathcal{I}$ if executing $\mathcal{B}_{obf}(i)$ produces a new sequence.
The objective is:
\begin{equation}
\max_{i \in \mathcal{I}} \left|\bigcup \text{N-grams}(\mathcal{B}_{obf}(i))\right|.
\end{equation}
We set $N=3$ to capture local interaction context while avoiding excessive state growth~\cite{xiao2025robust}.

Syscall boundaries serve as interaction anchors rather than the sole semantic source; computation between boundaries is represented through register/memory snapshots, branch traces, and observed state-transfer relations.

\noindent\textbf{\textit{Interaction Monitoring and Snapshot Generation.}}
During exploration, \toolname uses dynamic binary instrumentation (DBI)~\cite{nethercote2007valgrind} to monitor runtime side effects.
Rather than treating every executed instruction as equally important, the monitor listens for a predefined target syscall set $\Omega_{target}\subset\mathcal{S}_{syscall}$ covering critical file-system operations, e.g., \texttt{sys\_open}, \texttt{sys\_read}, and \texttt{sys\_write}, and network operations, e.g., \texttt{sys\_socket}, and \texttt{sys\_connect}. 

When a syscall in $\Omega_{target}$ is invoked, the DBI engine suspends execution and triggers a bounded capture handler.
The handler performs four steps:
(1) \textit{interaction identification}, reading the syscall number to obtain $\text{SysID}$;
(2) \textit{pointer tracing}, following architecture-specific calling conventions to inspect parameter registers and recursively trace memory pointers with depth limits and page-validity checks;
(3) \textit{payload extraction}, computing the payload length and safely dumping the referenced memory; and
(4) \textit{state construction}, recording the relevant registers and memory payloads as $\mathcal{R}_{init}$ and $\mathcal{M}_{init}$.
The resulting $B_{spec}$ serves as an execution-derived constraint for subsequent noise filtering, code synthesis, and behavioral validation.

\subsection{Specification-Driven Noise Filtering}
\label{sec:phase2}

This phase extracts trace-level instructions that contribute to the observed runtime behavior.
Since the input is a dynamic trace, it does not remove never-executed dead code; instead, it filters executed but behavior-irrelevant instructions, such as flattened dispatcher updates, opaque-predicate computations, and MBA junk computations that appear in the trace but do not affect the monitored interaction.

Given a dynamic trace $T=\langle i_1,i_2,\dots,i_n\rangle$ and a behavior snapshot $B_{spec}=\langle k,\text{SysID},\mathcal{R}_{init},\mathcal{M}_{init}\rangle$ from Phase I, where $i_k$ triggers the monitored interaction, this phase outputs a behavior-relevant instruction slice $L_{ext}\subseteq T$:
\begin{equation}
L_{ext}={i_j\in T \mid j\leq k \land \text{Path}(i_j\rightarrow i_k)\models B_{spec}},
\end{equation}
where $L_{ext}$ retains instructions in the transitive data-flow closure that contribute to the state specified by $B_{spec}$.

\noindent\textbf{\textit{Address Normalization.}}
Before slicing, \toolname normalizes runtime addresses to remove the offsets introduced by ASLR~\cite{snow2013just} and PIE~\cite{payer2012too}.
For each instruction $i_j$ with runtime address $Addr_{dyn}(i_j)$, the system parses runtime memory mappings and static ELF headers to identify the dynamic load base $Base_{dyn}$ and the static base $Base_{static}$.
It then computes the relocation offset $\Delta=Base_{dyn}-Base_{static}$ and maps the instruction to its static address:
\begin{equation}
Addr_{static}(i_j)=Addr_{dyn}(i_j)-\Delta .
\end{equation}
This mapping aligns dynamic traces with static binary locations, enabling address-consistent dependency analysis.

\noindent\textbf{\textit{Trace-Driven Backward Slicing.}}
Starting from the specification index $k$, \toolname traverses the trace backward and tracks an active dependency set $V$, initialized with $\mathcal{R}_{init}\cup\mathcal{M}_{init}$.
For each instruction $i_j$, the system computes its definition set $Def(i_j)$ and uses the set $Use(i_j)$.
If $Def(i_j)\cap V=\emptyset$, the instruction does not influence the monitored state and is discarded.
Otherwise, it is added to $L_{ext}$, and the dependency set is updated as $V\leftarrow (V\setminus Def(i_j))\cup Use(i_j)$.

To further remove control-flow-flattening noise, \toolname applies an orthogonality check.
If an instruction only updates scheduling-related architectural state, such as EFLAGS or RIP, and preserves the transitive dependency closure of $V$, it is treated as dispatcher noise and skipped.
By combining dependency closure filtering with this orthogonality check, \toolname extracts a compact slice $L_{ext}$ that captures the computation needed to reproduce the monitored interaction while excluding trace-level noise designed to mislead traditional deobfuscators.

Slicing is performed over executed instruction semantics rather than decompiler variables.
We recover register and memory uses and definitions from architecture-specific operand semantics, including implicit flags, condition codes, stack updates, and ABI syscall arguments.
For inter-procedural traces, traced callees are inlined, while unresolved indirect or library calls are summarized by observed arguments, return values, and memory effects.
Heap, global, and multi-syscall state are represented through monitored address ranges and ordered protocol windows.
Unresolved dependencies are retained conservatively, and unsupported side effects are marked unsupported.
The slice preserves observed-execution dependencies, but does not claim all-path static completeness.

\subsection{Specification-Guided Code Synthesis}

This phase lifts the behavior-relevant instruction slice $L_{ext}$ into readable source code that is consistent with the captured behavioral specifications. 
To reduce hallucinations and uncertainty in LLM-based code generation~\cite{liu2024lost, rawte2023survey}, we ground synthesis in factual execution constraints derived from dynamic runtime behavior.

The input consists of $L_{ext}$ and the behavior snapshot $B_{spec}=\langle k,\text{SysID},\mathcal{R}_{init},\mathcal{M}_{init}\rangle$; the output is candidate source code $\mathcal{C}_{candidate}$. 
We formulate synthesis as conditionally constrained generation:
\begin{equation}
\arg\max_{\mathcal{C}_{candidate}} P(\mathcal{C}_{candidate}\mid L_{ext}, B_{spec}).
\end{equation}

\noindent\textbf{\textit{Prompt Design.}}
The structured prompt guides the LLM to reason from objective runtime facts rather than unconstrained guesses. 
It contains four compact components.

First, the \textit{system prompt} assigns the model the role of an expert reverse engineer and defines the task as synthesizing readable code constrained by dynamic execution observations. 
Second, the \textit{input context} provides two factual inputs: the behavior-relevant instruction sequence $L_{ext}$ and the behavioral constraints derived from $B_{spec}$, including the initial state $S_{pre}$, composed of $\mathcal{R}_{init}$ and $\mathcal{M}_{init}$, and the target interaction boundary $S_{post}$, defined by the syscall identifier and target memory state. 
Third, the \textit{instruction body} asks the model to reason in three steps: infer the observed state-transfer relation $\Phi_{IO}:S_{pre}\rightarrow S_{post}$, abstract low-level instructions into high-level constructs, and construct standard C source code. 
Finally, the \textit{output constraint} requires the model to emit only the candidate source code, omitting explanations and markdown metadata.

By anchoring generation to the behavior-relevant slice and observed $S_{pre}\rightarrow S_{post}$ relation, this stage reduces reliance on flawed pseudocode and guides code toward the captured runtime behavior.

\subsection{Test-Based Behavioral Validation}
\label{sec:phase4}

This final phase tests whether the synthesized candidate code reproduces the monitored runtime behavior and observable side effects of the original binary on observed executions. 
It provides test-based behavioral validation with respect to captured specifications.

The input includes $\mathcal{C}_{candidate}$, $\mathcal{B}_{obf}$, and the Phase-I behavior snapshot $B_{spec}$; the output is $\mathcal{C}_{validated}$ if the candidate reproduces the captured behavior in a context-aware test harness.

\noindent\textbf{\textit{Context-Aware Test Harness.}}
\toolname compiles $\mathcal{C}_{candidate}$ into $Bin_{rec}$ and executes it in an isolated Unicorn-based emulator~\cite{unicorn}. 
The harness restores $\mathcal{R}_{init}$ and $\mathcal{M}_{init}$ from $B_{spec}$, aligning the candidate reconstruction with the monitored context of the original binary slice.

After synchronization, \toolname executes $\mathcal{B}_{obf}$ and $Bin_{rec}$ under the same observed context. 
At the target interaction point, it extracts $S_{orig}$ and $S_{rec}$ and checks behavioral consistency as:
\begin{equation} \begin{aligned} S_{orig} \equiv S_{rec} \iff\;& (\text{SysID}_{orig} = \text{SysID}_{rec}) \\ &\land\; (\text{Args}_{orig} = \text{Args}_{rec}) \\ &\land\; (\text{Mem}_{orig} = \text{Mem}_{rec}) . \end{aligned} \end{equation}
A candidate passes if the syscall identifiers, arguments, and relevant memory payloads match byte-level observations. 
This checks consistency with the captured snapshot, not all-input semantic equivalence.

\noindent\textbf{\textit{Discrepancy Feedback and Refinement.}}
If $S_{orig} \equiv S_{rec}$ holds on the validation cases, the candidate is promoted to $\mathcal{C}_{validated}$.
Otherwise, \toolname computes $\Delta_{diff}=S_{orig}\ominus S_{rec}$ and summarizes the mismatched syscall ID, argument, register, or memory payload as structured feedback to the LLM.
The synthesis engine then generates $\mathcal{C}_{candidate}^{(t+1)}$.
The loop stops when the monitored discrepancy is eliminated or $N_{max}$ is reached, whose sensitivity is studied in \S~\ref{sensitivity of N}.

\section{Experimental Setup}
\label{sec:exp_set}

\subsection{Research Questions}
We evaluate \toolname from five perspectives: effectiveness against baselines (RQ1), robustness across architectures and compiler optimizations (RQ2), contribution of individual components (RQ3), code quality and readability (RQ4), and downstream utility for CFG-based Linux malware detection (RQ5).

\subsection{Datasets}
We construct two datasets: a controlled \textit{Synthetic Benchmark} with available source-level tests, and \textit{MalBench}, a real-world malware benchmark for practical security evaluation.

\noindent\textbf{\textit{Synthetic Benchmark.}}
We construct the Synthetic Benchmark from CodeNet~\cite{puri2021codenet}.
From 341,069 C/C++ programs, we retain 1,573 programs that are non-trivial, executable, and suitable for controlled obfuscation, requiring average cyclomatic complexity greater than 15, more than 150 source lines, and successful compilation and evaluation-pipeline execution.
The retained set is therefore a controlled benchmark rather than a representative sample of all CodeNet programs. We obfuscate the retained programs using O-LLVM~\cite{OLLVM_Theory}, Hikari~\cite{hikari2017}, Tigress~\cite{tigress}, and Alcatraz~\cite{alcatraz2022}.
We apply six transformations, BCF, SUB, FLA, MBA, Opaque Predicates, and ImmMov, and combine them into 63 non-empty configurations from L1 to L6.
Each configuration is compiled across four architectures, x86, x64, ARM, and MIPS, and four optimization levels, O0--O3, with symbols and debugging information stripped.
This expansion yields 1,585,584 synthetic obfuscated binaries from the 1,573 retained source programs.
Appendix~\ref{app:synthetic} reports the filtering reason-code distribution and audit metadata.

\noindent\textbf{\textit{MalBench}.}
MalBench contains 500 active Linux obfuscated malware samples collected from VirusShare~\cite{virusshare} and MalwareBazaar~\cite{malwarebazaar} between 2023 and 2025. 
Following prior malware-evaluation practice~\cite{raff2017malware, patsakis2024assessing}, we retain samples flagged by at least 10 independent VirusTotal engines and exhibiting high cyclomatic complexity and heavily flattened control flow.
Because in-the-wild malware lacks source code and unit tests, we use observed behavioral consistency as the validation oracle: reconstructed code must reproduce the monitored syscall sequences and input-output interactions under the same input harness.
This oracle is bounded by observed executions and does not imply all-input semantic equivalence.
For downstream detection, MalBench uses malware-family-aware partitions and a benign corpus matched by platform, executable format, package source, and file-size range.
We retain hashes, family labels, first-seen dates, sandbox policy, and benign provenance as metadata, but do not redistribute live malware.
See Appendix~\ref{app:malbench} for more details about MalBench.

\begin{table}[t]
\centering
\caption{\textbf{Input-induced path coverage.}}
\label{tab:path_coverage}
\renewcommand{\arraystretch}{1.05}
\scriptsize
\resizebox{\columnwidth}{!}{%
\begin{tabular}{lcc}
\toprule
\textbf{Metric} & \textbf{Synthetic} & \textbf{MalBench} \\
\midrule

Basic-block coverage 
& $95\%$ 
& $88\%$ \\

Branch-edge coverage 
& $90\%$ 
& $70\%$ \\

Unique execution paths/sample 
& $12~[4,20]$ 
& $8~[2,16]$ \\

\bottomrule
\end{tabular}}
\end{table}

\begin{table*}[t]
\centering

\caption{\textbf{Effectiveness on the Synthetic Benchmark.}
CR and Pass@1 are reported for x64-O0 binaries across obfuscation levels L0--L6 as mean$_{\pm}$SD over five runs.}

\label{tab:rq1_dataset_a}
\renewcommand{\arraystretch}{1.15}
\resizebox{\textwidth}{!}{%
\begin{NiceTabular}{l|cc|cc|cc|cc|cc|cc|cc}
\CodeBefore
  \rowcolor[HTML]{F2F2F2}{4,6,8,10,12,14,16}
  \rowcolor[HTML]{FFF2CB}{18}
\Body
\hline
\multirow{2}{*}{\textbf{Method}} & \multicolumn{2}{c|}{\textbf{L0 (Baseline)}} & \multicolumn{2}{c|}{\textbf{L1}} & \multicolumn{2}{c|}{\textbf{L2}} & \multicolumn{2}{c|}{\textbf{L3}} & \multicolumn{2}{c|}{\textbf{L4}} & \multicolumn{2}{c|}{\textbf{L5}} & \multicolumn{2}{c}{\textbf{L6 (Extreme)}} \\ \cline{2-15} 
 & \textbf{CR} & \textbf{P@1} & \textbf{CR} & \textbf{P@1} & \textbf{CR} & \textbf{P@1} & \textbf{CR} & \textbf{P@1} & \textbf{CR} & \textbf{P@1} & \textbf{CR} & \textbf{P@1} & \textbf{CR} & \textbf{P@1} \\ \hline

Ghidra             & $25.4_{\pm 0.0}$ & $18.5_{\pm 0.0}$ & $12.1_{\pm 0.0}$ & $5.2_{\pm 0.0}$ & $4.5_{\pm 0.0}$ & $1.0_{\pm 0.0}$ & $1.2_{\pm 0.0}$ & $0.0_{\pm 0.0}$ & $0.0_{\pm 0.0}$ & $0.0_{\pm 0.0}$ & $0.0_{\pm 0.0}$ & $0.0_{\pm 0.0}$ & $0.0_{\pm 0.0}$ & $0.0_{\pm 0.0}$ \\

D810 & $30.5_{\pm 0.0}$ & $26.4_{\pm 0.0}$ & $25.2_{\pm 0.0}$ & $20.1_{\pm 0.0}$ & $15.4_{\pm 0.0}$ & $10.2_{\pm 0.0}$ & $5.1_{\pm 0.0}$ & $2.5_{\pm 0.0}$ & $1.8_{\pm 0.0}$ & $0.0_{\pm 0.0}$ & $0.0_{\pm 0.0}$ & $0.0_{\pm 0.0}$ & $0.0_{\pm 0.0}$ & $0.0_{\pm 0.0}$ \\

GooMBA         & $28.6_{\pm 0.0}$ & $20.1_{\pm 0.0}$ & $22.4_{\pm 0.0}$ & $15.3_{\pm 0.0}$ & $18.5_{\pm 0.0}$ & $10.4_{\pm 0.0}$ & $8.2_{\pm 0.0}$ & $4.1_{\pm 0.0}$ & $2.5_{\pm 0.0}$ & $0.0_{\pm 0.0}$ & $0.0_{\pm 0.0}$ & $0.0_{\pm 0.0}$ & $0.0_{\pm 0.0}$ & $0.0_{\pm 0.0}$ \\ \hline

ANGR & $45.2_{\pm 0.0}$ & $38.5_{\pm 0.0}$ & $35.1_{\pm 1.2}$ & $28.4_{\pm 1.4}$ & $22.4_{\pm 1.8}$ & $15.6_{\pm 2.0}$ & $10.5_{\pm 2.2}$ & $6.2_{\pm 2.4}$ & $4.2_{\pm 1.5}$ & $1.5_{\pm 0.8}$ & $1.0_{\pm 0.5}$ & $0.0_{\pm 0.0}$ & $0.0_{\pm 0.0}$ & $0.0_{\pm 0.0}$ \\

Syntia   & $50.5_{\pm 1.5}$ & $42.1_{\pm 1.8}$ & $38.4_{\pm 2.1}$ & $30.5_{\pm 2.4}$ & $28.6_{\pm 2.8}$ & $20.4_{\pm 3.1}$ & $18.2_{\pm 3.2}$ & $12.5_{\pm 3.5}$ & $10.1_{\pm 2.6}$ & $5.4_{\pm 2.0}$ & $4.5_{\pm 1.8}$ & $2.0_{\pm 1.2}$ & $1.2_{\pm 0.5}$ & $0.5_{\pm 0.2}$ \\ \hline

Llama-3.1 & $62.4_{\pm 1.2}$ & $55.2_{\pm 1.5}$ & $50.1_{\pm 1.8}$ & $42.5_{\pm 2.0}$ & $38.5_{\pm 2.4}$ & $30.1_{\pm 2.8}$ & $25.4_{\pm 3.1}$ & $18.5_{\pm 3.4}$ & $15.2_{\pm 2.8}$ & $10.4_{\pm 2.5}$ & $8.5_{\pm 2.1}$ & $5.2_{\pm 1.8}$ & $4.2_{\pm 1.5}$ & $2.1_{\pm 1.0}$ \\

Claude-3.5 & $68.5_{\pm 1.0}$ & $62.1_{\pm 1.2}$ & $58.4_{\pm 1.6}$ & $50.2_{\pm 2.0}$ & $45.2_{\pm 2.5}$ & $38.4_{\pm 2.8}$ & $30.5_{\pm 3.2}$ & $22.5_{\pm 3.6}$ & $20.1_{\pm 3.0}$ & $14.2_{\pm 2.8}$ & $12.4_{\pm 2.5}$ & $8.5_{\pm 2.2}$ & $6.5_{\pm 1.8}$ & $4.2_{\pm 1.5}$ \\

GPT-4o & $72.1_{\pm 0.8}$ & $66.5_{\pm 1.0}$ & $62.5_{\pm 1.4}$ & $55.4_{\pm 1.8}$ & $50.4_{\pm 2.2}$ & $42.1_{\pm 2.6}$ & $38.2_{\pm 3.0}$ & $28.5_{\pm 3.5}$ & $25.5_{\pm 3.2}$ & $18.4_{\pm 3.0}$ & $15.2_{\pm 2.8}$ & $10.5_{\pm 2.5}$ & $8.4_{\pm 2.1}$ & $5.5_{\pm 1.8}$ \\ \hline

CodeLlama   & $74.2_{\pm 0.9}$ & $68.1_{\pm 1.1}$ & $64.1_{\pm 1.5}$ & $56.2_{\pm 1.8}$ & $51.5_{\pm 2.4}$ & $44.1_{\pm 2.6}$ & $38.2_{\pm 3.1}$ & $31.5_{\pm 3.4}$ & $26.5_{\pm 3.2}$ & $18.2_{\pm 3.0}$ & $16.4_{\pm 2.8}$ & $10.5_{\pm 2.6}$ & $9.5_{\pm 2.4}$ & $5.2_{\pm 2.0}$ \\

Qwen2.5-C & $78.1_{\pm 0.7}$ & $72.2_{\pm 0.9}$ & $70.3_{\pm 1.2}$ & $64.1_{\pm 1.5}$ & $58.2_{\pm 2.0}$ & $51.2_{\pm 2.4}$ & $46.1_{\pm 2.8}$ & $38.5_{\pm 3.2}$ & $34.2_{\pm 3.0}$ & $26.4_{\pm 3.1}$ & $22.1_{\pm 2.8}$ & $14.2_{\pm 2.9}$ & $12.4_{\pm 2.6}$ & $7.5_{\pm 2.2}$ \\

DeepSeek-C& $81.5_{\pm 0.6}$ & $76.2_{\pm 0.8}$ & $74.1_{\pm 1.0}$ & $68.3_{\pm 1.2}$ & $64.2_{\pm 1.8}$ & $56.1_{\pm 2.1}$ & $51.4_{\pm 2.5}$ & $44.2_{\pm 2.8}$ & $38.2_{\pm 2.9}$ & $31.5_{\pm 3.2}$ & $26.4_{\pm 3.0}$ & $20.2_{\pm 3.1}$ & $16.5_{\pm 2.8}$ & $11.2_{\pm 2.6}$ \\ \hline

Recopilot & $84.2_{\pm 0.5}$ & $80.1_{\pm 0.6}$ & $78.5_{\pm 0.9}$ & $72.4_{\pm 1.2}$ & $68.5_{\pm 1.5}$ & $61.2_{\pm 1.8}$ & $56.2_{\pm 2.2}$ & $48.5_{\pm 2.5}$ & $44.1_{\pm 2.8}$ & $36.2_{\pm 3.0}$ & $31.2_{\pm 3.2}$ & $24.4_{\pm 3.4}$ & $20.5_{\pm 3.1}$ & $14.5_{\pm 2.9}$ \\

ChatDEOB   & $88.1_{\pm 0.4}$ & $84.2_{\pm 0.5}$ & $82.2_{\pm 0.8}$ & $76.1_{\pm 1.0}$ & $74.1_{\pm 1.2}$ & $68.2_{\pm 1.6}$ & $64.2_{\pm 2.0}$ & $56.4_{\pm 2.4}$ & $51.5_{\pm 2.6}$ & $44.1_{\pm 3.2}$ & $38.2_{\pm 3.5}$ & $31.5_{\pm 3.8}$ & $26.4_{\pm 3.6}$ & $20.2_{\pm 3.4}$ \\ \hline

LLM+Dyn & $88.3_{\pm 0.2}$ & $85.2_{\pm 0.3}$ & $83.5_{\pm 0.6}$ & $79.1_{\pm 1.5}$ & $77.1_{\pm 1.3}$ & $69.2_{\pm 1.4}$ & $66.8_{\pm 2.5}$ & $56.7_{\pm 2.2}$ & $56.6_{\pm 2.8}$ & $43.9_{\pm 2.7}$ & $44.2_{\pm 3.7}$ & $34.4_{\pm 3.2}$ & $35.1_{\pm 3.4}$ & $23.6_{\pm 3.5}$ \\ 

LLM+Dyn+Fb & $89.2_{\pm 0.5}$ & $86.1_{\pm 0.4}$ & $85.1_{\pm 0.3}$ & $80.5_{\pm 1.2}$ & $80.7_{\pm 0.9}$ & $71.2_{\pm 1.5}$ & $72.6_{\pm 2.6}$ & $61.9_{\pm 2.7}$ & $60.1_{\pm 3.3}$ & $51.2_{\pm 2.5}$ & $53.8_{\pm 2.8}$ & $42.5_{\pm 3.5}$ & $44.2_{\pm 2.8}$ & $39.1_{\pm 2.9}$ \\ \hline

\bftoolname & \textbf{94.5$_{\pm 0.2}$} & \textbf{92.1$_{\pm 0.3}$} & \textbf{93.2$_{\pm 0.3}$} & \textbf{90.5$_{\pm 0.4}$} & \textbf{91.5$_{\pm 0.4}$} & \textbf{88.2$_{\pm 0.5}$} & \textbf{89.4$_{\pm 0.5}$} & \textbf{84.5$_{\pm 0.6}$} & \textbf{87.2$_{\pm 0.6}$} & \textbf{81.1$_{\pm 0.8}$} & \textbf{84.5$_{\pm 0.8}$} & \textbf{77.2$_{\pm 1.0}$} & \textbf{81.4$_{\pm 1.1}$} & \textbf{74.5$_{\pm 1.2}$} \\ \hline
\end{NiceTabular}%
}
\end{table*}

\subsection{Evaluation Metrics}

We use three correctness-oriented metrics and three readability-oriented metrics. 
\textit{Compilation Rate (CR)} measures the percentage of reconstructed programs that compile without manual intervention. 
\textit{Unit Test Pass Rate (Pass@1)} measures whether the compiled reconstruction passes the available source-level tests in the Synthetic Benchmark. 
\textit{Execution Success Rate (ESR)} evaluates MalBench by checking whether the original and reconstructed binaries produce the same monitored syscall sequences and I/O interactions under the same input harness. 
For code quality, \textit{Cyclomatic Complexity Reduction (CCR)} measures control-flow simplification relative to raw IDA Pro decompilation, \textit{CodeBLEU} measures structural and data-flow similarity to available ground-truth source code, and \textit{Halstead Effort Reduction (HER)} measures lexical and cognitive-complexity reduction.

Rates are computed over the full evaluated denominator, including pass, fail, timeout, and unsupported outcomes.
LLM-based methods and \toolname are run five times with independent seeds and reported as mean$_{\pm}$SD, while deterministic tools are run once under fixed versions/settings and therefore have SD=0.0.

\noindent\textbf{\textit{Held-Out Behavioral Validation.}}
For dynamic specification experiments, we split observed executions into refinement and held-out sets before reconstruction.
The Synthetic Benchmark uses $16~[8,30]$ refinement inputs and $4~[2,10]$ held-out inputs per sample, while MalBench uses $12~[3,20]$ and $3~[2,8]$, reported as median [IQR].
Synthetic inputs come from source-level harnesses and generated tests, whereas MalBench inputs include sandbox stimuli such as arguments, staged files, environment settings, and replay configurations.
Only refinement inputs are used for prompts, feedback, and repair; held-out inputs are never exposed to any method or controlled variant.
We normalize or mask time-, PID-, path-, and network-dependent fields, and mark unreplayable executions as unsupported.
A reconstruction is credited only if it compiles and passes held-out source-level tests on the Synthetic Benchmark or syscall/I/O consistency checks on MalBench.
Thus, validation reduces trace-level overfitting but remains bounded by explored executions.

\noindent\textbf{\textit{Coverage Reporting.}} We report input-induced path coverage to make the scope of the observation-bounded oracle explicit.
Here, a sample denotes one reconstruction task, while an input denotes an execution stimulus, such as command-line arguments, stdin, staged files, environment settings, or sandbox replay configurations, used to trigger observable behavior.
Coverage is measured over the behavior-relevant target region extracted for each reconstruction task, rather than over the entire binary.
As shown in Table~\ref{tab:path_coverage}, the execution inputs cover 95\% of target-region basic blocks and 90\% of branch edges on the Synthetic Benchmark, and 88\% and 70\% on MalBench, respectively.
The median number of unique execution paths per sample is 12 on the Synthetic Benchmark and 8 on MalBench, indicating that validation is not limited to repeatedly replaying a single path.
These statistics do not imply complete path coverage or all-input semantic equivalence; instead, they make explicit the behavioral scope over which the reported reconstruction results are validated.

\subsection{Baselines}
To provide a comprehensive evaluation, we compare \toolname with representative baselines across static analysis, symbolic/dynamic analysis, hybrid synthesis, and LLM-based paradigms. 
We compare \toolname with representative baselines from four categories: static tools, including Ghidra~\cite{ghidra}, D810~\cite{d810github}, and GooMBA~\cite{goomba2021}; symbolic/dynamic tools, including ANGR~\cite{shoshitaishvili2016sok} and Syntia~\cite{blazytko2017syntia}; general-purpose and code LLMs, including GPT-4o~\cite{achiam2023gpt}, Claude-3.5-Sonnet~\cite{anthropic2024claude3}, Llama-3.1~\cite{llama31_2024}, Qwen2.5-Coder~\cite{hui2024qwen2}, CodeLlama~\cite{roziere2023code}, and DeepSeek-Coder~\cite{guo2024deepseek}; and domain-specific binary/deobfuscation models, including Recopilot~\cite{chen2025recopilot} and ChatDEOB~\cite{choi2024chatdeob}. 
To separate method effects from information-budget effects, we additionally include two controlled LLM variants: LLM+DynEvidence (LLM+Dyn), which receives the same bounded runtime traces and monitored snapshots as \toolname, and LLM+DynEvidence+Feedback (LLM+Dyn+Fb), which further receives the same compile/test feedback budget but does not perform trace-driven slicing or behavior-specification-guided constraint construction.

Because LLM-based approaches lack native support for direct end-to-end recovery from raw binaries, we first utilized standard decompilers to translate the obfuscated inputs into pseudocode before feeding them into the LLMs.
See Appendix~\ref{app:baselines} for more details about baselines.

\subsection{Implementation Details}
All experiments were conducted on a workstation equipped with dual Intel Xeon Gold 6248R processors, 512GB of RAM, and four NVIDIA RTX 4090 GPUs, running Ubuntu 22.04 LTS. All LLM-based baselines, controlled variants, and \toolname use fixed prompt templates, temperature $=0.2$, top-$p=1.0$, a 4K output cap, and the input budgets specified in Appendix~\ref{app:baselines}.
We record exact model identifiers, provider/checkpoint information, prompt-template hashes, retry limits, context-truncation policy, and seeds in the artifact logs.
Held-out inputs are excluded from all prompts, feedback messages, repair rounds, and truncation decisions.
More implementation details can be found in Appendix~\ref{implementation}.

\section{Evaluation}
\subsection{RQ1: Effectiveness}

Table~\ref{tab:rq1_dataset_a} reports CR and Pass@1 on x64-O0 Synthetic Benchmark binaries, with model names abbreviated for space. Specifically, we draw the following key observations.

\noindent\textbf{\textit{Baselines degrade under compositional obfuscation.}}
Existing methods degrade as obfuscation becomes more compositional.
Traditional static tools, such as Ghidra, D810, and GooMBA, and symbolic or dynamic techniques, such as ANGR and Syntia, drop to near-zero accuracy at higher levels because rule-based recovery and bounded exploration struggle with flattened control flow, artificial branches, and complex predicates.
LLM-based methods are more resilient but still degrade sharply: GPT-4o drops from 66.5\% Pass@1 at L0 to 5.5\% at L6, while the domain-specific ChatDEOB reaches only 20.2\% Pass@1 at L6.
It suggests that larger models or task-specific fine-tuning are insufficient when the input is dominated by misleading obfuscated structure.

\noindent\textbf{\textit{\bfittoolname preserves effectiveness under severe obfuscation.}}
In contrast, \toolname maintains substantially higher effectiveness under severe obfuscation.
At L6, \toolname achieves 74.5\% Pass@1, outperforming the strongest evaluated baseline, ChatDEOB, by 54.3 percentage points.
It also achieves the highest compilation rate at L6, 81.4\%, indicating that behavior-constrained synthesis and closed-loop feedback help reduce invalid code generation.
These results support the central design choice of reconstructing code from behavior-relevant slices and observed runtime constraints rather than directly transforming the obfuscated code structure.

\noindent\textbf{Controlled variants isolate the method effect.}
To test whether the gains of \toolname come from the proposed pipeline rather than from simply giving an LLM more runtime evidence, Table~\ref{tab:rq1_dataset_a} includes two evidence-matched controlled variants.
\textit{LLM+DynEvidence} receives the same syscall traces and monitored I/O snapshots as \toolname, and \textit{LLM+DynEvidence+Feedback} further receives the same compile and test feedback budget.
However, these variants do not perform trace-driven slicing or behavior-specification-guided constraint construction, and held-out inputs are never exposed during refinement.
Although the controlled variants improve over prior LLM baselines, the strongest one still trails \toolname by 22.2 percentage points in average Pass@1 and by 35.4 percentage points at L6.
This result indicates that the improvement is not explained by a larger information budget alone, but by how \toolname organizes runtime observations into behavior-relevant slices and synthesis constraints.

\begin{table}[t]
\centering
\caption{\textbf{Execution success rate on MalBench.}}
\label{tab:malbench_esr}
\renewcommand{\arraystretch}{0.95}
\setlength{\tabcolsep}{4pt}
\begin{tabular}{lr@{\hspace{12pt}}|@{\hspace{12pt}}lr}
\toprule
\textbf{Method} & \textbf{ESR (\%)} & \textbf{Method} & \textbf{ESR (\%)} \\
\midrule
\rowcolor[HTML]{F2F2F2}
Ghidra        & 0.0  & GPT-4o              & 15.2 \\
D810          & 0.0  & CodeLlama           & 18.4 \\
\rowcolor[HTML]{F2F2F2}
GooMBA        & 0.0  & Qwen2.5-C           & 22.5 \\
ANGR          & 4.2  & DeepSeek-C          & 25.4 \\
\rowcolor[HTML]{F2F2F2}
Syntia        & 6.5  & Recopilot           & 31.2 \\
Llama-3.1     & 8.4  & ChatDEOB            & 36.5 \\
\rowcolor[HTML]{F2F2F2}
Claude-3.5    & 12.5 & \cellcolor[HTML]{F2F2F2}\bftoolname & \cellcolor[HTML]{F2F2F2}\textbf{65.8} \\
\bottomrule
\end{tabular}
\end{table}

\noindent\textbf{\textit{\bfittoolname remains effective on real-world malware.}}
Table~\ref{tab:malbench_esr} reports Execution Success Rate (ESR) on MalBench, a dataset of real-world Linux obfuscated malware.
In this setting, traditional static tools such as Ghidra and D810 obtain zero ESR, while ANGR and Syntia reach only 4.2\% and 6.5\%, respectively.
LLM-based methods show limited robustness, ranging from 15.2\% for GPT-4o to 36.5\% for ChatDEOB.
By contrast, \toolname achieves 65.8\% ESR, outperforming the strongest baseline by 29.3 percentage points.
This result suggests that behavior-specification-guided reconstruction remains useful beyond controlled synthetic obfuscation, especially when real malware combines structural noise, environmental dependencies, and anti-analysis behaviors.
We interpret this result as behavioral consistency over the family-aware MalBench split, not as recovery of complete malware source code.

\begin{figure*}[!t]
  \centering
  \includegraphics[width=1\linewidth]{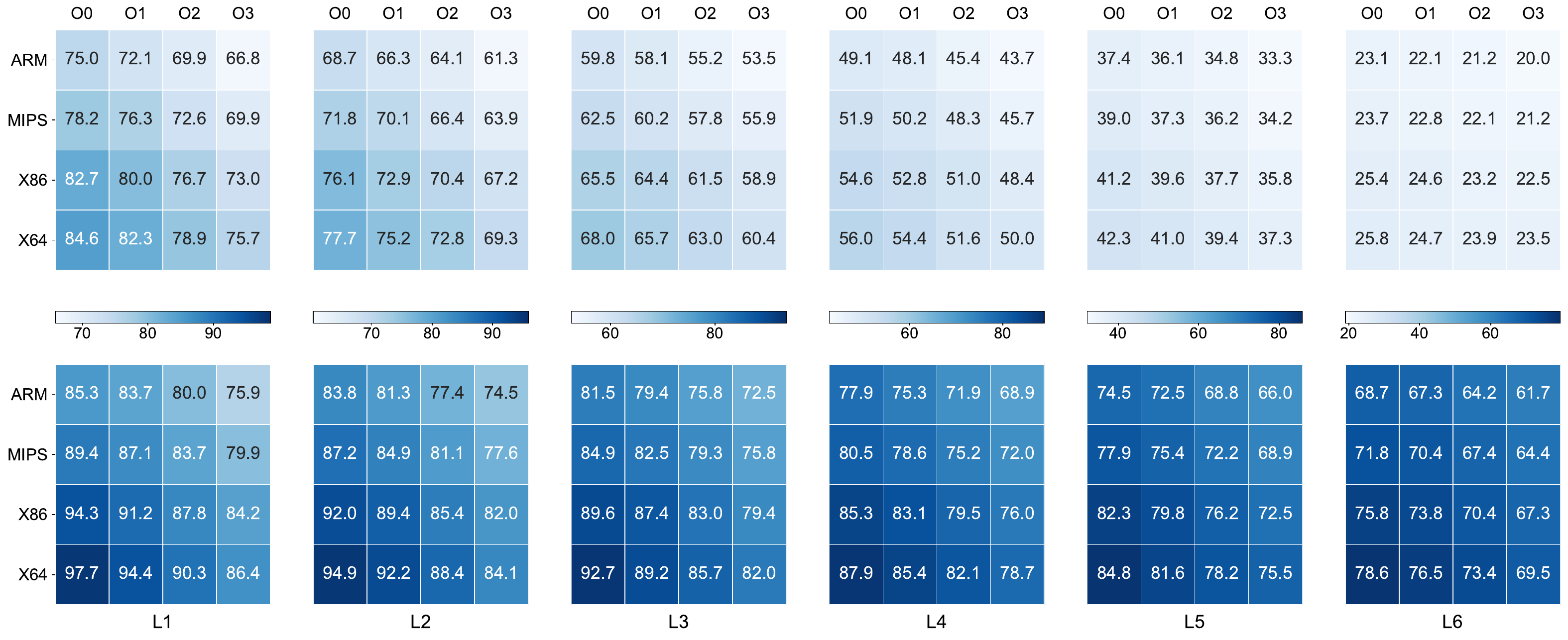} 
  \caption{Performance (Pass@1) across architectures and optimizations. Rows 1 and 2 show results for ChatDEOB~\cite{choi2024chatdeob} and \toolname across six obfuscation levels, respectively.}
  \label{fig:RQ2.}
\end{figure*}

\noindent\textbf{\textit{\bfittoolname covers many cases missed by baselines.}}
\begin{table}[t]
\centering
\caption{\textbf{Sample-level overlap analysis.}}
\label{tab:overlap_failure}
\resizebox{0.38\textwidth}{!}{%
\begin{tabular}{lrr}
\toprule
\textbf{Category} & \textbf{\#Cases} & \textbf{Percentage} \\
\midrule

Both Succeed & 1,037 & 34.6\% \\
\rowcolor[HTML]{F2F2F2}
\toolname-Only & 1,331 & 44.4\% \\

Baseline-Only & 82 & 2.7\% \\
\rowcolor[HTML]{F2F2F2}
Both Fail & 550 & 18.3\% \\
\bottomrule
\end{tabular}}
\end{table}

We analyze sample-level overlap on 3,000 Synthetic Benchmark cases randomly sampled from L2, L4, and L6. D810, Syntia, and ChatDEOB represent static, dynamic/symbolic, and learning-based baselines, respectively. A sample is counted as a baseline success if any of the three baselines produces a reconstruction that passes the corresponding test suite.
Table~\ref{tab:overlap_failure} shows that \toolname succeeds alone on 1,331 cases, or 44.4\%, indicating that it recovers many samples missed by representative baselines. These \toolname-only cases mainly involve control-flow flattening, opaque predicates, MBA transformations, or decompiler-induced type and structure loss, where static structure becomes unreliable. Baseline-only cases account for 82 cases, or 2.7\%, and mainly involve local expression-level obfuscation or simple opaque predicates that static or symbolic methods can simplify. The both-fail category accounts for 18.3\%, showing that some cases remain difficult for all evaluated approaches, especially when generated inputs do not cover the relevant behavior or synthesis cannot produce compilable code under the captured constraints. Overall, the overlap analysis shows that \toolname expands the set of recoverable cases rather than merely improving already-solved ones, while still leaving room for complementary static and symbolic techniques on localized simplification tasks.

\begin{table*}[t]
\centering
\caption{\textbf{Ablation study on the Synthetic Benchmark.}
CR and Pass@1 are reported across obfuscation levels L0--L6 after removing each core component from \toolname.}
\label{tab:ablation_study}
\renewcommand{\arraystretch}{1.2}
\resizebox{\textwidth}{!}{%
\begin{NiceTabular}{l|cc|cc|cc|cc|cc|cc|cc}
\CodeBefore
  \rowcolor[HTML]{F2F2F2}{4,6}
\Body
\hline
\multirow{2}{*}{\textbf{Configuration}} 
& \multicolumn{2}{c|}{\textbf{L0 (No Obf.)}} 
& \multicolumn{2}{c|}{\textbf{L1}} 
& \multicolumn{2}{c|}{\textbf{L2}} 
& \multicolumn{2}{c|}{\textbf{L3}} 
& \multicolumn{2}{c|}{\textbf{L4}} 
& \multicolumn{2}{c|}{\textbf{L5}} 
& \multicolumn{2}{c}{\textbf{L6 (Extreme)}} \\ 
\cline{2-15} 
& \textbf{CR} & \textbf{Pass@1} 
& \textbf{CR} & \textbf{Pass@1} 
& \textbf{CR} & \textbf{Pass@1} 
& \textbf{CR} & \textbf{Pass@1} 
& \textbf{CR} & \textbf{Pass@1} 
& \textbf{CR} & \textbf{Pass@1} 
& \textbf{CR} & \textbf{Pass@1} \\ 
\hline

$w/o$ Backward Slicing 
& 88.4 & 82.5 
& 82.0 & 73.1 
& 76.5 & 65.4 
& 68.2 & 54.8 
& 60.2 & 45.1 
& 51.5 & 36.2 
& 42.5 & 28.5 \\

$w/o$ I/O Constraints 
& 91.5 & 85.2 
& 86.8 & 80.4 
& 82.1 & 74.5 
& 77.5 & 67.2 
& 72.4 & 61.2 
& 67.8 & 54.5 
& 62.5 & 48.4 \\

$w/o$ Iterative Refinement 
& 92.1 & 88.4 
& 89.5 & 85.2 
& 87.4 & 82.1 
& 84.2 & 77.5 
& 80.5 & 71.4 
& 76.8 & 65.8 
& 72.6 & 60.5 \\ 
\hline

\textbf{\bftoolname{} (Full)} 
& \textbf{94.5} & \textbf{92.1} 
& \textbf{93.2} & \textbf{90.5} 
& \textbf{91.5} & \textbf{88.2} 
& \textbf{89.4} & \textbf{84.5} 
& \textbf{87.2} & \textbf{81.1} 
& \textbf{84.5} & \textbf{77.2} 
& \textbf{81.4} & \textbf{74.5} \\ 
\hline
\end{NiceTabular}
}
\end{table*}

\subsection{RQ2: Robustness}
To evaluate the robustness across architectures and optimizations, Figure~\ref{fig:RQ2.} shows the results across different instruction set architectures, including ARM, MIPS, X86, and X64, and compilation optimization levels from O0 to O3.
We compare \toolname with ChatDEOB, the strongest evaluated baseline in RQ1.

\noindent\textbf{Architectures.} 
ChatDEOB shows a clear performance drop across different architectures.
At Level-1 under O0, ChatDEOB achieves 84.86\% on X64 but drops to 74.69\% on ARM.
This gap suggests a limitation in its reliance on decompiled pseudocode and architecture-specific structural patterns.
It performs better on prevalent X64 structures but struggles with the distinct control-flow and instruction representations found in RISC architectures.
In contrast, \toolname demonstrates stronger cross-architecture robustness.
At Level-1 under O0, \toolname achieves 97.58\% on X64 and maintains 86.05\% on ARM.
Even under extreme L6 obfuscation, \toolname maintains a 78.57\% success rate on X64-O0 and 68.92\% on ARM-O0, outperforming ChatDEOB's 25.51\% and 22.90\% in the same settings.
By relying on dynamic execution observations, \toolname reduces the syntax gap between different instruction sets.

\noindent\textbf{Optimizations.} 
We observe consistent performance degradation across compilation optimization levels.
Higher optimizations (O2 and O3) introduce aggressive register allocation, instruction reordering, and function inlining, which can disrupt the expected pseudocode layout.
ChatDEOB is sensitive to these changes.
On the X64 architecture at L1, its performance decreases from 84.86\% under O0 to 75.57\% under O3.
Under severe L6 obfuscation, the combination of aggressive optimization and complex transformations causes ChatDEOB to drop to 23.00\% on X64-O3 and 20.82\% on ARM-O3.
In contrast, \toolname is less affected by compiler-induced structural changes.
At Level-1 on X64, \toolname retains an 86.45\% success rate even at O3.
Under the most demanding L6 configuration, \toolname achieves 69.94\% on X64-O3 and 61.82\% on ARM-O3.
These results suggest that behavioral constraints help \toolname maintain recovery effectiveness across architecture and optimization differences, although performance still decreases as optimization and obfuscation complexity increase.

\subsection{RQ3: Ablation Study}

To assess component contributions, we conduct ablation studies on three variants of \toolname, each removing one key component: trace-driven backward slicing (\textit{$w/o$ Backward Slicing}), dynamic I/O constraints during synthesis (\textit{$w/o$ I/O Constraints}), and context-aware differential testing with closed-loop feedback (\textit{$w/o$ Iterative Refinement}). 
As shown in Table~\ref{tab:ablation_study}, we evaluate all variants on the Synthetic Benchmark under x64-O0 across obfuscation levels L0--L6. Removing backward slicing and directly feeding raw traces to the LLM substantially degrades performance, especially under extreme obfuscation. 
At L6, Pass@1 drops from 74.5\% to 28.5\%, indicating that slicing is crucial for filtering dispatcher logic or MBA junk instructions before synthesis. 
Removing dynamic I/O constraints also reduces L6 Pass@1 to 48.4\%, showing that behavioral specifications are important for grounding LLM generation. 
Finally, without iterative refinement, L6 Pass@1 falls to 60.5\%, suggesting that differential testing and closed-loop feedback improve test-based behavioral consistency.

\begin{table*}[t]
\centering
\caption{\textbf{Code quality and readability on the Synthetic Benchmark.}}
\label{tab:rq4_code_quality}
\renewcommand{\arraystretch}{1.15}
\resizebox{\textwidth}{!}{%
\begin{NiceTabular}{l|cccccc|cccccc|cccccc}
\CodeBefore
  \rowcolor[HTML]{F2F2F2}{4,6} 
  \rowcolor[HTML]{FFF2CB}{8}   
\Body
\hline
\multirow{2}{*}{\textbf{Method}} 
& \multicolumn{6}{c|}{\textbf{CodeBLEU ($\uparrow$)}} 
& \multicolumn{6}{c|}{\textbf{CCR \% ($\uparrow$)}} 
& \multicolumn{6}{c}{\textbf{HER \% ($\uparrow$)}} \\ 
\cline{2-19} 
& \textbf{L1} & \textbf{L2} & \textbf{L3} & \textbf{L4} & \textbf{L5} & \textbf{L6} 
& \textbf{L1} & \textbf{L2} & \textbf{L3} & \textbf{L4} & \textbf{L5} & \textbf{L6} 
& \textbf{L1} & \textbf{L2} & \textbf{L3} & \textbf{L4} & \textbf{L5} & \textbf{L6} \\ 
\hline

D810~\cite{d810github}           
& $36.4$ & $25.1$ & $12.5$ & $8.4$  & $6.1$  & $5.2$  
& $31.5$ & $18.5$ & $8.2$  & $4.5$  & $2.8$  & $2.1$  
& $34.2$ & $21.0$ & $10.5$ & $3.2$  & $-2.1$ & $-5.8$ \\

Syntia~\cite{blazytko2017syntia} 
& $25.4$ & $16.8$ & $10.2$ & $7.1$  & $5.2$  & $4.5$  
& $22.5$ & $14.1$ & $8.4$  & $5.0$  & $3.1$  & $2.1$  
& $18.6$ & $10.4$ & $5.2$  & $3.1$  & $2.0$  & $1.5$  \\

GPT-4o~\cite{achiam2023gpt}       
& $55.4$ & $46.2$ & $38.5$ & $29.4$ & $22.1$ & $18.5$ 
& $58.2$ & $45.1$ & $35.4$ & $26.8$ & $19.5$ & $15.2$ 
& $52.1$ & $41.5$ & $32.4$ & $23.1$ & $16.8$ & $12.4$ \\

DeepSeek-Coder~\cite{guo2024deepseek} 
& $62.5$ & $54.1$ & $45.6$ & $36.2$ & $29.5$ & $24.5$ 
& $65.2$ & $56.4$ & $48.5$ & $37.1$ & $26.4$ & $20.2$ 
& $60.4$ & $51.2$ & $42.1$ & $31.5$ & $23.4$ & $18.5$ \\

ChatDEOB~\cite{choi2024chatdeob}  
& $71.2$ & $63.8$ & $55.4$ & $47.1$ & $39.8$ & $34.5$ 
& $74.5$ & $66.1$ & $58.2$ & $46.5$ & $36.2$ & $28.4$ 
& $71.8$ & $64.2$ & $56.5$ & $45.1$ & $34.8$ & $25.6$ \\ 
\hline

\textbf{\bftoolname{} (Ours)} 
& \textbf{88.5} & \textbf{84.6} & \textbf{80.2} & \textbf{75.8} & \textbf{72.1} & \textbf{68.4} 
& \textbf{93.5} & \textbf{89.2} & \textbf{85.4} & \textbf{82.1} & \textbf{78.8} & \textbf{76.2} 
& \textbf{91.2} & \textbf{87.1} & \textbf{82.6} & \textbf{78.4} & \textbf{75.2} & \textbf{72.8} \\ 
\hline
\end{NiceTabular}%
}
\end{table*}
\begin{table*}[t]
  \centering
  \caption{\textbf{Malware detection performance on obfuscated binaries.}}
  \label{tab:downstream_detection}
  \renewcommand{\arraystretch}{1.2} 
  \resizebox{0.85\textwidth}{!}{
  \begin{NiceTabular}{lcccc}
      \CodeBefore
      \rowcolor[HTML]{F2F2F2}{3,5} 
    \Body
    \toprule
    \textbf{Detection Pipeline} & \textbf{Precision (\%)} & \textbf{Recall (\%)} & \textbf{F1-Score (\%)} & \textbf{Accuracy (\%)} \\
    \midrule
    HawkEye-Linux & 61.4 & 48.2 & 54.1 & 58.2 \\
    \midrule
    HawkEye-Linux + D810 & 64.2  (2.8$\uparrow$) & 56.0  (7.8$\uparrow$) & 59.8  (5.7$\uparrow$) & 62.4  (4.2$\uparrow$) \\
    HawkEye-Linux + ChatDEOB & 76.8  (15.4$\uparrow$) & 72.4  (24.2$\uparrow$) & 74.5  (20.4$\uparrow$) & 75.8  (17.6$\uparrow$) \\
    \textbf{HawkEye-Linux + \bftoolname{} (Ours)} & \textbf{92.4  (31.0$\uparrow$)} & \textbf{90.1  (41.9$\uparrow$)} & \textbf{91.2  (37.1$\uparrow$)} & \textbf{91.5  (33.3$\uparrow$)} \\
    \bottomrule
  \end{NiceTabular}}
\end{table*}

\subsection{RQ4: Code Quality and Readability}
Table~\ref{tab:rq4_code_quality} presents the results on the Synthetic Benchmark using CodeBLEU, Cyclomatic Complexity Reduction (CCR), and Halstead Effort Reduction (HER) to evaluate structural complexity and human readability of the source-level reconstructions across escalating obfuscation levels.
Traditional static and dynamic tools exhibit rapid degradation as obfuscation intensifies, with D810 dropping to a negative HER of $-5.8\%$ at Level-6.
This indicates that rule-based matching can generate excessive \texttt{goto} statements and redundant intermediate variables, making the decompiled output more complex and harder to read. Similarly, LLM-based methods experience a performance drop under compound obfuscations, as seen with the domain-specific ChatDEOB decreasing from 74.5\% CCR at Level-1 to 28.4\% CCR and 25.6\% HER at Level-6.
This vulnerability suggests that relying primarily on static syntactic patterns or supervised fine-tuning leaves models susceptible to hallucination when multi-layered structural noise disrupts the expected pseudocode distribution.

In contrast, \toolname maintains stronger performance across all evaluated metrics and obfuscation levels.
At Level-1, it achieves 88.5\% CodeBLEU and reduces cyclomatic complexity by 93.5\%.
Under Level-6, it sustains 68.4\% CodeBLEU, 76.2\% CCR, and 72.8\% HER, outperforming ChatDEOB by large margins in complexity reduction.
By anchoring source-level reconstruction to dynamic execution observations rather than static parsing alone, \toolname avoids many injected fake branches and redundant predicates, producing more compact and readable observed-behavior-consistent code.

\subsection{RQ5: Downstream Utility}

To evaluate whether \toolname improves downstream security analysis, we conduct a malware detection experiment on heavily obfuscated Linux executables. We use HawkEye-Linux, a HawkEye-style CFG-based detector adapted to Linux ELF artifacts~\cite{xu2021hawkeye}. The detector extracts CFGs from executable artifacts, represents basic blocks with instruction-level features, learns graph-level program embeddings with a graph neural network, and performs malware/benign classification with a fixed classifier head. It does not directly consume reconstructed source code, pseudocode, ASTs, or manually engineered semantic features. 

\noindent\textbf{\textit{All pipelines are normalized through executable artifacts.}}
The direct binary-analysis condition uses the original obfuscated ELF binary as the detector artifact.
For deobfuscation-enhanced conditions, each front-end first produces a simplified or reconstructed representation, which is then compiled into a normalized Linux ELF artifact using the same compiler configuration, linker setting, wrapper template, and stripping policy whenever possible.
HawkEye-Linux then extracts CFGs and instruction-level features from the resulting executable artifact.
If artifact construction fails within the timeout, we fall back to the original obfuscated ELF and record the case as an artifact-construction failure.
All pipelines use the same family-aware train/validation/test split, benign-matching policy, graph-extraction procedure, model architecture, hyperparameters, random seed, and threshold-selection rule.
Thus, RQ5 measures whether different deobfuscation front-ends produce executable artifacts whose graph representations are more useful for downstream Linux malware detection.

\noindent\textbf{\textit{The evaluation uses matched malware and benign corpora.}}
We construct a balanced corpus with 500 in-the-wild Linux malware samples from MalBench and 500 benign Linux executables collected from standard Ubuntu repositories.
The malware set covers active families such as Mirai, Gafgyt, and ransomware samples, and includes binaries protected by multi-layered anti-analysis transformations.
Both malware and benign samples are processed under the same construction and graph-extraction protocol to reduce label leakage.

\begin{table*}[t]
\centering
\caption{\textbf{Impact of the iteration limit ($N_{max}$) on the Synthetic Benchmark under the L6 configuration.}}
\label{tab:iteration_ceiling_transposed}
\renewcommand{\arraystretch}{1.2}
\resizebox{0.75\textwidth}{!}{%
\begin{NiceTabular}{lcccccccc}
\CodeBefore
  \rowcolors{2,4}{gray!10}{white}
\Body
\toprule
\textbf{Iteration $N_{max}$} & \textbf{1} & \textbf{2} & \textbf{3} & \textbf{4} & \textbf{5} & \textbf{6} & \textbf{8} & \textbf{10} \\ \midrule
\textbf{Pass@1 (\%) }      & 60.5\%     & 67.8\%     & 71.4\%     & 73.5\%     & 74.5\%     & 74.7\%     & 74.8\%     & 74.8\%      \\
\textbf{Gain ($\Delta$ \%)}  & -          & +7.3\%     & +3.6\%     & +2.1\%     & +1.0\%     & +0.2\%     & +0.1\%     & +0.0\%      \\
\textbf{Avg. Time (s)}     & 5.2        & 10.4       & 15.7       & 21.1       & 26.5       & 32.1       & 43.5       & 55.2        \\ \bottomrule
\end{NiceTabular}%
}
\end{table*}
\begin{table*}[t]
\centering
\caption{\textbf{Effect of slice length under L6 obfuscation.}
Longer extracted slices $|L_{ext}|$ are associated with lower synthesis accuracy.}
\label{tab:slice_length}
\resizebox{0.75\textwidth}{!}{%
\begin{NiceTabular}{cccl}
\CodeBefore
  \rowcolors{3}{white}{gray!10}
\Body
\toprule
\textbf{Slice Length}      & \textbf{Proportion of} & \textbf{Cumulative Pass@1} & \textbf{Primary}            \\
\textbf{($|L_{ext}|$)}     & \textbf{Dataset (\%)}  & \textbf{(Success Rate \%)} & \textbf{Failure Mode}       \\ \midrule
$\le 50$ instructions      & 25.4\%                 & 92.4\%                     & Minor Syntax Errors         \\
$51 - 100$ instructions    & 34.6\%                 & 85.1\%                     & Variable Type Mismatch      \\
$101 - 250$ instructions   & 24.5\%                 & 71.3\%                     & Logic Inversion             \\
$251 - 500$ instructions   & 10.5\%                 & 45.2\%                     & Missing Dependencies        \\
$> 500$ instructions       & 5.0\%                  & 21.6\%                     & Context Truncation          \\ \bottomrule
\end{NiceTabular}%
}
\end{table*}

\noindent\textbf{\textit{\bfittoolname improves downstream malware detection.}}
Table~\ref{tab:downstream_detection} reports the detection results across the evaluated pipelines.
When HawkEye-Linux analyzes the original obfuscated binaries directly, it achieves only 58.2\% accuracy and 54.1\% F1-score, indicating that aggressive structural obfuscation weakens CFG-based malware classification.
Adding deobfuscation front-ends improves performance to different degrees.
D810 provides limited gains, increasing accuracy by 4.2 percentage points and F1-score by 5.7 percentage points, while ChatDEOB improves accuracy by 17.6 percentage points and F1-score by 20.4 percentage points.
The \toolname-enhanced pipeline achieves the best result, with 91.5\% accuracy and 91.2\% F1-score, improving over direct binary analysis by 33.3 and 37.1, respectively.

\noindent\textbf{\textit{The improvement comes from behavior-consistent artifacts.}}
D810 often produces limited gains because static pattern matching struggles with combinatorial obfuscation and may leave fragmented control-flow artifacts.
ChatDEOB can recover some high-level structure, but under heavy obfuscation, it may generate logic that is inconsistent with observed execution behavior, which can mislead the downstream detector.
In contrast, \toolname uses runtime behavior constraints during reconstruction, reducing the impact of injected structural noise and producing normalized executable artifacts that better preserve behaviorally relevant control-flow patterns.

\section{Discussion}

\subsection{Sensitivity and Convergence of the Closed-Loop Iteration (\texorpdfstring{$N_{max}$}{N\_max})}
\label{sensitivity of N}

An important hyperparameter in the \toolname pipeline is the maximum number of iterations ($N_{max}$) during the test-based behavioral validation stage.
This parameter controls how many times the LLM can refine the candidate source code based on differential behavior feedback ($\Delta_{diff}$).
To evaluate the trade-off between test-based recovery effectiveness and computational overhead, we conducted a sensitivity analysis on $N_{max}$ using the Synthetic Benchmark under the L6 obfuscation configuration.
Table~\ref{tab:iteration_ceiling_transposed} reports the Unit Test Pass Rate (Pass@1), the cumulative time overhead per slice, and the incremental improvement at each iteration step.

\noindent\textbf{\textit{Optimal Iteration Threshold.}}
Relying solely on synthesis without refinement ($N_{max}=1$) yields a Pass@1 of 60.5\%.
In the second iteration, the success rate increases by 7.3\%, indicating that differential feedback helps resolve syntactic errors and local behavioral deviations.
As iterations continue, however, the success rate plateaus at 74.5\% after $N_{max}=5$, while the average time overhead continues to grow and reaches 55.2 seconds at $N_{max}=10$.
These results indicate diminishing returns in later iterations.
Therefore, we set $N_{max}=5$ as the default threshold to balance test-based recovery effectiveness and computational efficiency.

\noindent\textbf{\textit{Analysis of Unresolved Cases.}}
We further analyzed the samples that remained unsuccessful after the 5th iteration and identified two main limiting factors.
First, \textit{Incomplete Behavior-Relevant Slices}: If the initial backward data-flow slice ($L_{ext}$) misses implicit dependencies, for example, due to pointer aliasing beyond the monitored memory boundary, the LLM may not infer the missing logic regardless of the provided refinement prompts.
Second, \textit{Context Window Exceedance}: For particularly large slices, repeatedly appending behavioral discrepancy logs ($\Delta_{diff}$) can exceed the LLM's effective context capacity.
This causes the model to lose track of earlier constraints, sometimes resulting in oscillation between incorrect candidate programs across iterations.

\noindent\textbf{\textit{Implications for LLM-based Synthesis.}}
These findings highlight the practical role of LLMs in our setting: they are more effective as \textit{constraint-guided generators} than as standalone deobfuscation solvers.
The feedback mechanism provides useful test-based signals, but it still relies on the completeness of the behavior-relevant slice extracted in Phase II and the coverage of observed executions captured in Phase I.
Consequently, future work should focus on improving dynamic taint tracking, pointer-alias modeling, and memory-constraint capture, rather than simply increasing the LLM iteration limit.

\subsection{Impact of Instruction Slice Length and Context Boundaries}

To evaluate the boundary of the LLM's semantic comprehension, we analyzed the synthesis success rate, measured by Pass@1, across different slice-length ($|L_{ext}|$) intervals using the Synthetic Benchmark under the L6 configuration.
Table~\ref{tab:slice_length} presents the distribution of slice lengths, the corresponding success rates, and the primary failure modes observed during synthesis.
It is worth noting that the 150-LOC threshold in dataset construction refers to the original source-level programs/functions, whereas $|L_{ext}|$ measures the behavior-relevant instruction slice extracted for a specific observed behavior after Phase II.
Therefore, a non-trivial source program can still yield a relatively short slice when only a small portion of the program contributes to a monitored interaction.

Specifically, \toolname maintains high Pass@1 values of 92.4\% and 85.1\% for short-to-medium slices ($\leq 100$ instructions), which constitute 60\% of the dataset.
For these lengths, failures primarily involve minor syntactic issues or variable type mismatches that can often be corrected by closed-loop refinement.
However, as the slice length exceeds 250 instructions, the success rate drops to 45.2\%, and further declines to 21.6\% for slices longer than 500 instructions.

The empirical results indicate that the model's ability to maintain low-level data-flow context degrades when processing long, dense assembly sequences.
Unlike high-level source code, assembly instructions lack explicit structural boundaries and exhibit tightly coupled register and memory dependencies.
As a result, the LLM may omit dependencies, truncate output, or generate inconsistent control/data-flow structures over extended sequences.
To process slices exceeding 500 instructions, future extensions of \toolname should investigate hierarchical chunking strategies.
For example, recursively partitioning large dependency graphs into smaller behavior-preserving sub-tasks before LLM inference may help the system maintain higher translation quality for large computational functions.

\subsection{Threats to Validity}

\subsubsection{Internal Validity}

\noindent\textbf{\textit{Emulation, Instrumentation, and Slicing Accuracy.}}
Our validation relies on execution states captured through CPU emulation and dynamic binary instrumentation.
Bugs in QEMU/DBI, missed implicit effects, or architecture-specific memory operations may make the captured state incomplete and cause false positives.
Similarly, trace-driven backward slicing may miss behavior-relevant semantics when syscall memory boundaries, pointer aliases, or dynamic address mappings are imprecise.
We mitigate these risks by validating syscall IDs, arguments, and memory payloads at monitored interaction points, conservatively over-approximating memory boundaries according to architecture-specific ABIs~\cite{matz2013system}, and using dynamic address mapping for runtime pointer references.
Nevertheless, our validation remains bounded by captured executions and does not prove all-input semantic equivalence.

\noindent\textbf{\textit{LLM Non-Determinism.}}
Semantic lifting may be affected by LLM non-determinism, prompt sensitivity, backend updates, and refinement trajectories.
We reduce this effect by using fixed prompt templates, low-temperature decoding, bounded feedback budgets, and five independent runs with mean$_{\pm SD}$ reporting.
Held-out inputs are never exposed to prompts, repair messages, or refinement feedback.

\noindent\textbf{\textit{Sparse Observable Behavior.}}
\toolname is effective when inputs trigger observable interaction boundaries and behavior-relevant state transitions.
CPU-local or dormant computations that do not affect monitored registers, memory regions, return values, or syscall/I/O behavior may provide weak specifications and limit reconstruction.
Our path-coverage reporting makes this scope explicit, but uncovered behavior remains outside the observation-bounded oracle.

\subsubsection{External Validity}

\noindent\textbf{\textit{Architectural and OS Scope.}}
Our implementation targets Linux binaries and evaluates four architectures: x86, x64, ARM, and MIPS.
Generalizing \toolname to Windows, macOS, or bare-metal firmware requires additional engineering because observable interaction boundaries may differ from POSIX syscalls.
For example, Windows binaries may require instrumentation at Native API or user-mode library boundaries rather than direct syscall interfaces.
Thus, while the behavior-specification-guided synthesis paradigm is not inherently POSIX-specific, the current implementation is Linux-oriented.

\noindent\textbf{\textit{Benchmark Selection and Contamination.}}
The Synthetic Benchmark is derived from CodeNet, a public dataset that may overlap with LLM pretraining data.
We therefore treat benchmark contamination as a validity threat and do not claim complete decontamination.
We mitigate this risk by removing duplicate source hashes, assigning train/evaluation partitions at the problem level before architecture and obfuscation expansion, withholding ground-truth source code and held-out tests from prompts and feedback, and evaluating correctness through compiled held-out executions rather than textual similarity alone.
The retained programs are also selected for executability, test availability, and obfuscatability, so they should be interpreted as a controlled benchmark rather than a representative sample of all CodeNet programs.

\section{Conclusion}
In this paper, we presented \toolname, a behavior-specification-guided synthesis approach for binary deobfuscation.
\toolname uses dynamic execution traces and interaction snapshots to constrain source-level reconstruction.
This design helps mitigate semantic loss introduced by compilation and decompilation, and enables the recovered code to be validated against observed runtime behavior.
Our evaluation shows that \toolname improves robustness under heavy obfuscation, produces more analyzable reconstructions, and provides useful executable artifacts for downstream malware analysis.
These results suggest that behavior-constrained synthesis offers a practical direction for binary deobfuscation, especially when static structure and decompiler output are unreliable.


\bibliographystyle{IEEEtran}
\bibliography{IEEEabrv,references}

\newpage
\appendices

\section{Synthetic Benchmark Audit}
\label{app:synthetic}
Table~\ref{tab:synthetic_filtering} reports the exclusion reason for each filtered CodeNet program.
Each excluded program is assigned one reason according to the first failed filtering stage, making the categories mutually exclusive.
The retained 1,573 programs are executable, testable, and obfuscatable, but are not intended to represent the full CodeNet distribution.

\begin{table}[h!]
\centering
\caption{\textbf{Synthetic Benchmark filtering reason codes.}}
\label{tab:synthetic_filtering}
\renewcommand{\arraystretch}{1.05}
\scriptsize
\resizebox{\columnwidth}{!}{%
\begin{tabular}{lrr}
\toprule
\textbf{Filtering outcome} & \textbf{\# Programs} & \textbf{Fraction} \\
\midrule

Initial C/C++ programs 
& 341,069 
& 100.00\% \\

\midrule
Low complexity or short source 
& 238,748
& 70\% \\

Compilation or evaluation-pipeline failure 
& 51,165 
& 15\% \\

Unavailable or flaky tests 
& 27,285 
& 8\% \\

Unsupported dependencies or external resources 
& 10,237 
& 3\% \\

Obfuscator failure 
& 5,114 
& 1.5\% \\

Timeout-triggered exclusion 
& 2,386 
& 0.7\% \\

Duplicate or leakage-control removal 
& 1,025
& 0.3\% \\

Other filtering failures 
& 3,536
& 1.04\% \\

\midrule
Retained source programs 
& 1,573 
& 0.46\% \\

\bottomrule
\end{tabular}}
\end{table}

For each retained program, we record the CodeNet problem ID, source hash, language mode, compiler version, compilation flags, test metadata, random seed, and partition ID.
The full obfuscation expansion applies four obfuscators, six transformations, four architectures, and four optimization levels, yielding $1{,}573 \times 63 \times 4 \times 4 = 1{,}585{,}584$ binaries.

\section{MalBench and Downstream Detection Details}
\label{app:malbench}

MalBench is handled as sensitive security data.
We release hashes and audit metadata sufficient to reproduce the selection and evaluation protocol, but do not redistribute live malware binaries.
MalBench contains 500 Linux malware samples collected from VirusShare and MalwareBazaar during 2023--2025.
A sample is retained only if it is a Linux executable, is flagged by at least 10 VirusTotal engines, and is not an exact duplicate by SHA-256 hash.
Corrupted, non-Linux, non-executable, and duplicate samples are removed.

For downstream detection, we construct a matched benign corpus of 500 Linux executables from standard Ubuntu repositories.
Benign samples are matched by architecture, executable format, and file-size bucket, using the ranges $<$100KB, 100KB--1MB, 1--5MB, 5--20MB, and $>$20MB, to reduce trivial corpus differences.
We use a family-aware 70/10/20 train/validation/test split with split seed 42, assigning all samples from the same malware family to the same split.
Held-out test samples are never used for prompt construction, refinement, detector training, or threshold selection. All malware executions are isolated in Docker/QEMU with restricted network and filesystem policies.
Outbound network traffic is disabled or redirected to a sink, the root filesystem is read-only, writable state is confined to a \textit{tmpfs} workspace, syscall and I/O logging are enabled, and each execution is bounded by a 30-minute timeout.
For RQ5, all pipelines use the same HawkEye-Linux detector configuration, graph-extraction procedure, model architecture, hyperparameters, training seed, validation split, and threshold-selection rule.
The threshold is selected on the validation split and fixed before test evaluation.
We compute 95\% confidence intervals with 10,000 bootstrap resamples using bootstrap seed 42.

\section{Baseline Information Budgets}
\label{app:baselines}

To avoid conflating method design with available evidence, we assign each baseline an explicit information budget.
Static-only tools receive only the raw binary or tool-specific IR, with no LLM prompt, dynamic trace, behavioral snapshot, repair, or feedback.
Decompiler+LLM and LLM+SamePrompt receive only IDA/Ghidra pseudocode under the same prompt format, decoding setting, 16K/4K input-output token budget, and at most one compile-error repair round.
LLM+Dyn uses the same prompt but additionally receives the bounded syscall-trace summary and monitored snapshot fields $\langle k,\text{SysID},\mathcal{R}_{init},\mathcal{M}_{init}\rangle$, with a 32K/4K token budget.
LLM+Dyn+Fb further receives up to $N_{\max}=5$ compile/test feedback rounds generated only from the refinement split.
In contrast, \toolname receives behavior-relevant slices, monitored snapshots, behavior-specification-guided constraints, and up to $N_{\max}=5$ refinement rounds.

All LLM-based variants use temperature $=0.2$ and top-$p=1.0$.
Dynamic evidence, when enabled, is shared in the same serialized format across controlled variants and \toolname.
Feedback rounds are generated only from refinement inputs; held-out inputs are never used for prompt construction, repair, feedback generation, or iterative refinement.
Thus, LLM+Dyn and LLM+Dyn+Fb test whether the gains in Table~\ref{tab:rq1_dataset_a} can be explained by stronger runtime evidence or feedback alone, while \toolname additionally tests the effect of trace-driven slicing and behavior-specification-guided constraint construction.

\section{Implementation Details}
\label{implementation}

\begin{table*}[t]
\centering
\caption{\textbf{Prompt template used by \toolname.}}
\label{tab:prompt_template}
\renewcommand{\arraystretch}{1.08}
\scriptsize
\resizebox{\textwidth}{!}{%
\begin{tabular}{p{2.6cm}p{13.2cm}}
\toprule
\textbf{Prompt Part} & \textbf{Template Content} \\
\midrule

System prompt 
&
You are an expert reverse engineer and systems programmer. 
Your task is to synthesize readable, compilable C code that is consistent with the provided behavior-relevant binary slice and monitored runtime observations. 
Do not invent behavior that is not supported by the given slice, register/memory state, syscall boundary, or I/O constraints. 
Prefer simple and portable C constructs. \\

\midrule

Input context 
&
\textbf{Target:} Reconstruct the behavior of the following binary slice as C code. \newline
\textbf{Architecture/ABI:} \texttt{<ARCH\_ABI>} \newline
\textbf{Required function signature:} \texttt{<FUNC\_SIGNATURE>} \newline
\textbf{Behavior-relevant instruction sequence} $L_{ext}$: \newline
\texttt{<SLICE\_INSTRUCTIONS>} \newline
\textbf{Initial runtime state} $S_{pre}$: \newline
Registers $\mathcal{R}_{init}$: \texttt{<REGISTER\_STATE>} \newline
Memory $\mathcal{M}_{init}$: \texttt{<MEMORY\_STATE>} \newline
\textbf{Target interaction boundary} $S_{post}$: \newline
Syscall identifier: \texttt{<SYSID>} \newline
Syscall arguments / target buffers: \texttt{<SYSCALL\_ARGS\_AND\_BUFFERS>} \newline
Expected target memory or I/O effect: \texttt{<EXPECTED\_EFFECT>} \newline
\textbf{Compilation constraints:} \texttt{<COMPILATION\_CONSTRAINTS>} \\

\midrule

Instruction body 
&
Perform the following reasoning internally before writing the code. \newline
1. Infer the observed state-transfer relation $\Phi_{IO}:S_{pre}\rightarrow S_{post}$ from the slice and runtime observations. \newline
2. Abstract low-level register, memory, pointer, arithmetic, and branch operations into high-level C constructs. \newline
3. Construct a self-contained C implementation of \texttt{<FUNC\_SIGNATURE>} that preserves the observed behavior at the target interaction boundary. \newline
The generated code should compile under \texttt{<COMPILER\_AND\_FLAGS>} and should not rely on undefined behavior, undeclared external state, or unavailable libraries unless explicitly listed in the constraints. \\

\midrule

Output constraint 
&
Output only the candidate C source code for \texttt{<FUNC\_SIGNATURE>}. 
Do not include explanations, markdown fences, comments about the task, confidence scores, or intermediate reasoning. 
If a helper function is necessary, include it in the same output. \\

\bottomrule
\end{tabular}}
\end{table*}

\subsection{Execution Environment and Baseline Settings}

All experiments were conducted on a workstation equipped with dual Intel Xeon Gold 6248R processors, 512GB of RAM, and four NVIDIA RTX 4090 GPUs, running Ubuntu 22.04 LTS.
Malware execution was isolated using Docker and QEMU.
For traditional static deobfuscators, we directly exported their generated pseudocode or recovered IR for evaluation.
For dynamic deobfuscators, ANGR~\cite{shoshitaishvili2016sok} used a Depth-First Search \texttt{SimulationManager} with a 30-minute timeout, while Syntia~\cite{blazytko2017syntia} used Monte Carlo Tree Search with 50,000 iterations and 100 input-output samples.
For LLM-based baselines, we provided IDA-generated pseudocode with fixed system prompts to standardize the semantic lifting task.

\begin{table}[t]
\centering
\caption{\textbf{Evaluation scale and resource summary.}}
\label{tab:scale_cost}
\renewcommand{\arraystretch}{1.05}
\scriptsize
\resizebox{\columnwidth}{!}{%
\begin{tabular}{lr}
\toprule
\textbf{Item} & \textbf{Value} \\
\midrule
Unique source programs & 1,573 \\
Expanded synthetic binaries & 1,585,584 \\
RQ1 evaluated subset & x64-O0, 100,672 binaries (incl. L0) \\
RQ2 architecture/optimization sweep & 1,585,584 binaries \\
MalBench samples & 500 \\
Completed non-timeout runs & 94.2\% \\
Timeout runs & 3.9\% \\
Unsupported runs & 1.9\% \\
Median runtime/sample & 8.5 min [4.2,17.6] \\
Median LLM calls/sample & 3 [2,5] \\
Total compute & 72K CPU-h / 1.2B LLM tokens \\
\bottomrule
\end{tabular}}
\end{table}

\subsection{LLM Protocol and Prompt Template}

All LLM-based baselines, controlled variants, and \toolname use fixed prompt templates, temperature $=0.2$, top-$p=1.0$, and a 4K output-token cap.
Pseudocode-only variants use a 16K/4K input-output budget, while evidence-augmented variants and \toolname use a 32K/4K budget to accommodate trace and snapshot summaries.
Decompiler+LLM and LLM+SamePrompt allow at most one compile-error repair round; LLM+Dyn+Fb and \toolname allow up to $N_{\max}=5$ refinement rounds using only refinement-split discrepancies.
The evaluated models include GPT-4o-2024-08-06, Claude-3.5-Sonnet-20241022, Llama-3.1-70B-Instruct, Qwen2.5-Coder-32B-Instruct, CodeLlama-34B-Instruct, and DeepSeek-Coder-33B-Instruct.
Closed models are queried through their official APIs, and open models are evaluated from fixed local checkpoints.
Exact model identifiers, provider or checkpoint information, seeds, prompt-template hashes, serialized evidence files, and request logs are recorded in the artifact.

Held-out inputs are never included in prompts, repair messages, feedback, truncation decisions, or refinement traces. When the context exceeds the input budget, we preserve function signatures, call sites, monitored snapshot fields, and the most recent behavior-relevant trace window, and truncate non-executed decompiler text first. Table~\ref{tab:prompt_template} shows the fixed prompt template used by \toolname. Baseline and controlled variants use the same output format and decoding constraints.

\subsection{Evaluation Scale and Resource Usage}

Table~\ref{tab:scale_cost} summarizes the evaluation scale and resource usage.
The 1,585,584 synthetic binaries denote the full obfuscated benchmark expansion, rather than the denominator of every experiment: RQ1 evaluates the x64-O0 subset including L0, RQ2 evaluates the architecture/optimization sweep, and MalBench uses 500 malware samples under family-aware splits.
All reported rates are computed over the corresponding evaluated denominator, where completed non-timeout runs include both pass and fail validation outcomes, and timeout or unsupported runs are not silently removed.
Runtime is reported as median [IQR] per evaluated sample, and LLM calls are counted over synthesis-invoking runs.
CPU time includes dataset construction, binary execution, baseline execution, slicing, synthesis orchestration, and validation.
LLM tokens include both prompt and generated tokens across LLM-based baselines, controlled variants, and \toolname.

\end{document}